\documentclass[a4paper,12pt,english]{article}
\usepackage[utf8]{inputenc}
\usepackage{latexsym}
\usepackage{microtype}
\usepackage{amsmath}
\usepackage{nicefrac}

\usepackage{tabularx}
\usepackage{colortbl}
\usepackage{hhline}
\usepackage{rotfloat}
\usepackage{soul}
\usepackage{amssymb,amsfonts,amsbsy}
\usepackage{xcolor}
\usepackage{graphicx}
\usepackage{float}
\usepackage[caption = false]{subfig}
\usepackage[english]{babel}
\usepackage{collcell}
\usepackage{multirow}
\usepackage{pgf}
\usepackage{natbib}
\usepackage{setspace}
\usepackage[a4paper,top=3cm,bottom=3cm,left=2.5cm,right=2.5cm,marginparwidth=1.75cm]{geometry}

\begin{document}
\title{Clustering of bivariate satellite time series: a  quantile  approach}
\author{Victor Muthama Musau \thanks{Address for correspondence: Victor Muthama Musau,
		Department of Pure and Applied Sciences Kirinyaga University, KENYA. 
		E-mail: \texttt{vmusau@kyu.ac.ke}. }\\
		Department of Pure and Applied Sciences, \\Kirinyaga University, Kenya \\
 Carlo Gaetan \\ 
 Dipartimento di Scienze Ambientali, Informatica e Statistica,\\ Universit\`a  Ca' Foscari - Venezia, Italy\\
 Paolo Girardi \\ Dipartimento di Scienze Ambientali, Informatica e Statistica,\\ Universit\`a  Ca' Foscari - Venezia, Italy\\
}
\date{\today}
\maketitle
\begin{abstract}
	Clustering has received much attention in Statistics and Machine learning with the aim of developing statistical models and autonomous algorithms which are capable of acquiring information from raw data in order to perform exploratory analysis.  Several techniques have been developed to cluster sampled univariate vectors only considering the average value over the whole period and as such they have not been able to explore fully the underlying distribution as well as other features of the data, especially in presence of structured time series.
	We propose a model-based clustering technique that is based on quantile regression permitting us to cluster bivariate time series at different quantile levels. We model the within cluster density using asymmetric Laplace distribution allowing us to take into account asymmetry in the distribution of the data. We evaluate the performance of the proposed technique through a simulation study. The method is then applied to cluster time series observed from Glob-colour satellite data related to trophic status indices with aim of evaluating their temporal dynamics  in order to identify homogeneous areas, in terms of trophic status, in the Gulf of Gabes.
\end{abstract}
\newpage
\section{Introduction}\label{sec1}

The European Community with the European Union Water Framework Directive 2000/60/EC (WFD) indicated a series of trophic status indicators with the scope to monitor the status of the sea-water in order to restore and protect water-bodies from further degradation \citep{directive2000european,Alikas:2015}. 
The light diffuse attenuation coefficient at 490 nm (KD-490) is an ecologically important water property that
provides information about the availability of light to underwater communities  which influences ecological
processes and biogeochemical cycles in natural waters \citep{Yang:2020}.
Together with KD-490, the Chlorophyll type-a (Chl-a) is commonly used as a proxy for phytoplankton biomass and as an indicator for eutrophication; high concentration levels may lead to hypoxic or anoxic events while low levels may result in improvement in water quality \citep{Dabuleviciene:2020}. The joint use of two complementary indices as the KD-490 and the Chl-a allows to take into account different aspects of water quality ranging from availability of light to underwater communities to the trophic status.

In this context, the classification of areas with a different level of impact may help Institutions to define a program of conservation and environmental protection. The European WFD 2000/60/EC set a series of rules to classify areas at different level of ``impact'' considering summary statistics (average, geometric mean, or percentile) of a single indicator over a predetermined temporal window (1 year, 5 years, etc, \ldots) with respect to a reference condition represented by no or very low human pressure \citep{poikane2010defining}. The definition of ``reference condition'' may be particularly complex and limited to each specific context \citep{Pardo:2012}. 

Measures of the KD-490 and Chl-a indices are obtained  from satellite sensors.
The use  of satellite data needs to consider aspects relative to unreliability for different reasons and mainly due to the presence of cloud contamination or malfunctioning of the sensor onboard the satellite; pre-processing procedures are often required to remove some site and measurements before using the data for the application \citep{alvera2012outlier,Stafoggia:2017}.

In this paper, we concentrate our efforts to overcome the classification based on simple summary statistics considering the temporal component in order to provide more reliable results since it takes into account the time dynamics of a water bodies indicator (i.e. seasonal behaviour, inter-annual variability, etc\ldots). 
In fact, standard clustering techniques 
were often applied considering the summary statistics of the determinants of interest, and so, potentially valuable information about the temporal behaviours (e.g. trend, peaks, and seasonal patterns) is lost.

In the environmental field time series clustering has gained popularity for grouping time series with similar temporal pattern 
covering a wide series of applications and approaches \citep{Cazelles:2008,Giraldo:2012,shi2013remote,Finazzi:2015,Haggarty:2015,Gaetan:2017}.

 Moreover very often the summary statistics suffer from the lack of robustness in presence of contaminated data.
In this respect quantile regression appears attractive due to the possibility to overcome these pre-processing issues with the possibility of choosing a particular quantile of interest \citep{Barbosa:2011, Monteiro:2012}. 
Another juncture concerns that most of the published work considered the clustering of univariate response variable \citep{Barbosa:2011,Monteiro:2012} or even a different weight between average value and temporal trend \citep{li2016bivariate} while some other attempts were performed taking into account the joint distribution of two or more variables of interest or the joint modelling of more quantiles. \citep{gaetan2016clustering,Zhang:2019,Sottile:2019}.

Therefore we propose a Bayesian clustering technique to define groups of temporal patterns that are similar considering the quantile of interest on bivariate time series. More in detail, the application regards monthly time series of Chl-a concentrations and KD-490 levels, obtained by satellite data sensors over the Gulf of Gabes (Tunisia), a Mediterranean zone with important biological resources and rich coastal, marine, and freshwater ecosystems. Since the last few decades, due to fast and uncontrolled urbanization and industrialization, the Gulf of Gabes is experiencing an irreversible degradation of the local coastal area. \citep{ayadi:2015,el:2017}. 

The paper is organized as follows. The next section illustrates our data on Chl-a and KD490 on the Gulf of the Gabes. In Section 3 we introduce the Bayesian quantile regression with an extension to the bivariate case and proposing our clustering procedure. In Section 4 we present a simulation study for illustrating the performances and peculiarities of the procedure. Section 5 reports the results of the application to the Chlorophyll-a and KD-490 satellite data on the Gulf of the Gabes. In the last section  we discuss the relative strengths and weaknesses of our proposal.

\section{Chl-a and KD-490 levels in the Gulf of Gabes}\label{sec2}
The Gulf of Gabes is a Mediterranean zone with important biological resources and rich coastal, marine and freshwater ecosystems. In the last decades the Gulf of Gabes reported several environmental problems mainly due to the presence of human activities associated with over-fishing and seabed trawling while the presence of chemical factories in the Sfax site resulting on a wide wastewater pollution \citep{Aloulou:2012,Rabaoui:2013,Zaghden:2014,Fourati:2018}.
The results had led  to  several issues as the local appearance of red tides \citep{hamza:1994} as well as the changes or decline of the distribution of some marine species \citep{el:2016,el:2018}.

The European Community indicates  the diffuse Chlorophyll type-a (Chl-a) concentrations as a trophic status indicator of the sea-water. High Chl-a levels may lead to hypoxic or anoxic events.  KD-490 indicates how the solar light can penetrate to deeper water and can be used to evaluate the potential disturbance to the water ecosystem.

In this work, we considered two datasets comprising the Chl-a concentrations and the KD-490 levels made available by ACRI (\texttt{hermes.acri.fr}) in the framework of the GlobColour Project (\texttt{www.globcolour.info}).
The datasets were formed by monthly values of Chl-a concentrations and KD-490 levels in the Gulf of Gabes from January, 2003 to December, 2011 for a total of $108$ time-points. Data were obtained by calibrating Ocean Colour data provided by different satellite missions, such as MERIS, SeaWiFS and MODIS. For each month, gridded data with 1.8-km resolution are available and  a  grid of $4,033$ points  covers the entire Gulf of Gabes.
\begin{figure}
	\centering
	\begin{tabular}{c}
\includegraphics[width=0.95\linewidth]{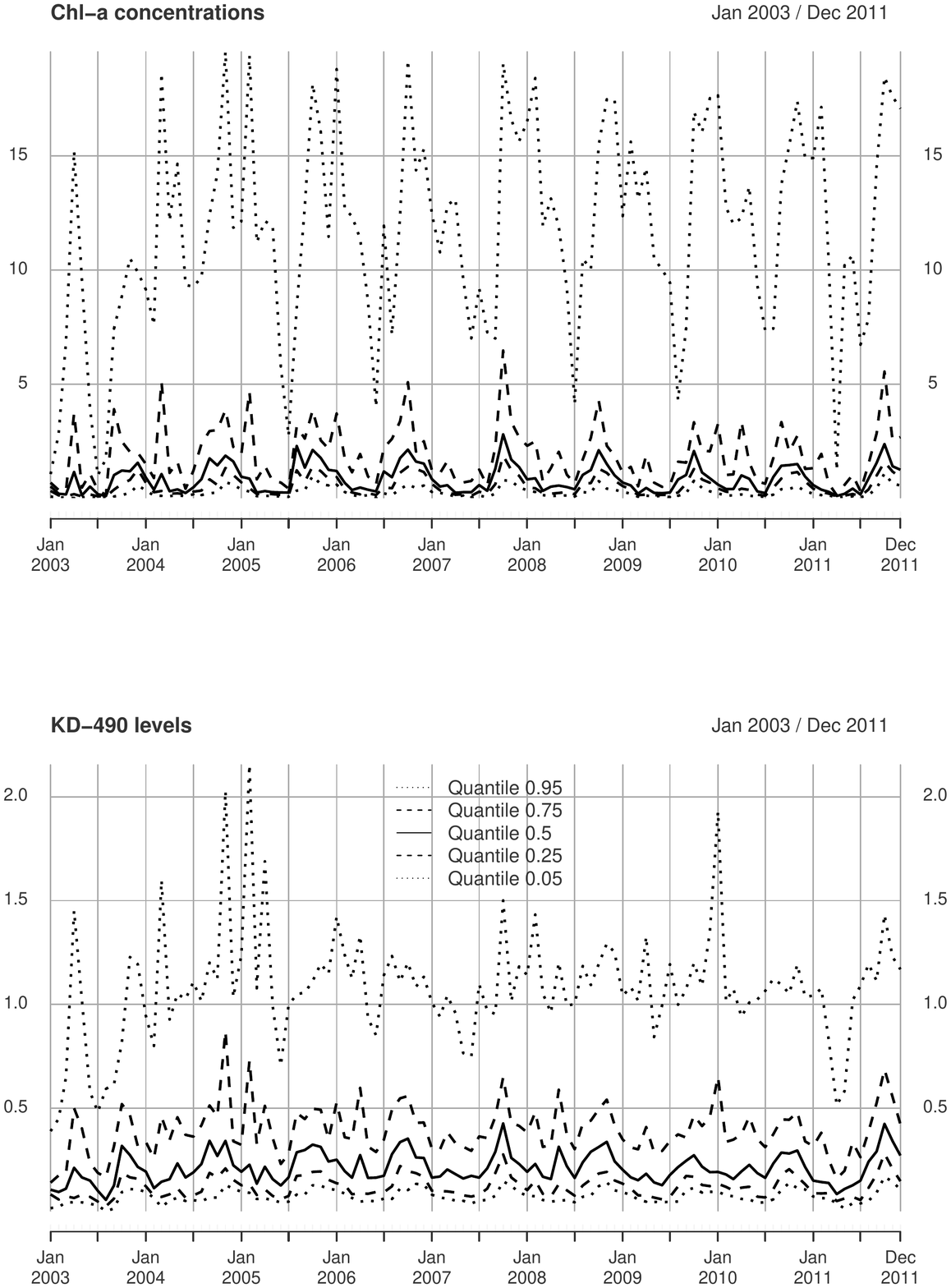} 
	\end{tabular}
	\caption{\textcolor{black}{Temporal profile for the quantiles (0.05, 0.25, 0.50, 0.75, and 0.95). For each time we calculated the empirical quantiles over the 4033 time-series.}}\label{fig:2plus}
\end{figure}
\textcolor{black}{As reported in Figure \ref{fig:2plus}, both Chl-a concentrations and KD-490 levels exhibit a seasonal cyclical pattern, more evident in the case of chlorophyll, with the presence of a peak during the spring period. While the KD-490 is related to the sea water turbidity more or less correlated to environmental factors (heavy rain, wind direction, etc\dots), the Chl-a index reflects the seasonal bloom in vegetation, directly connected to the seasonal variation of the sunlight window and the sea water temperature.}

\begin{figure}
	\centering
	\begin{tabular}{cc}
	\includegraphics[width=0.95\linewidth]{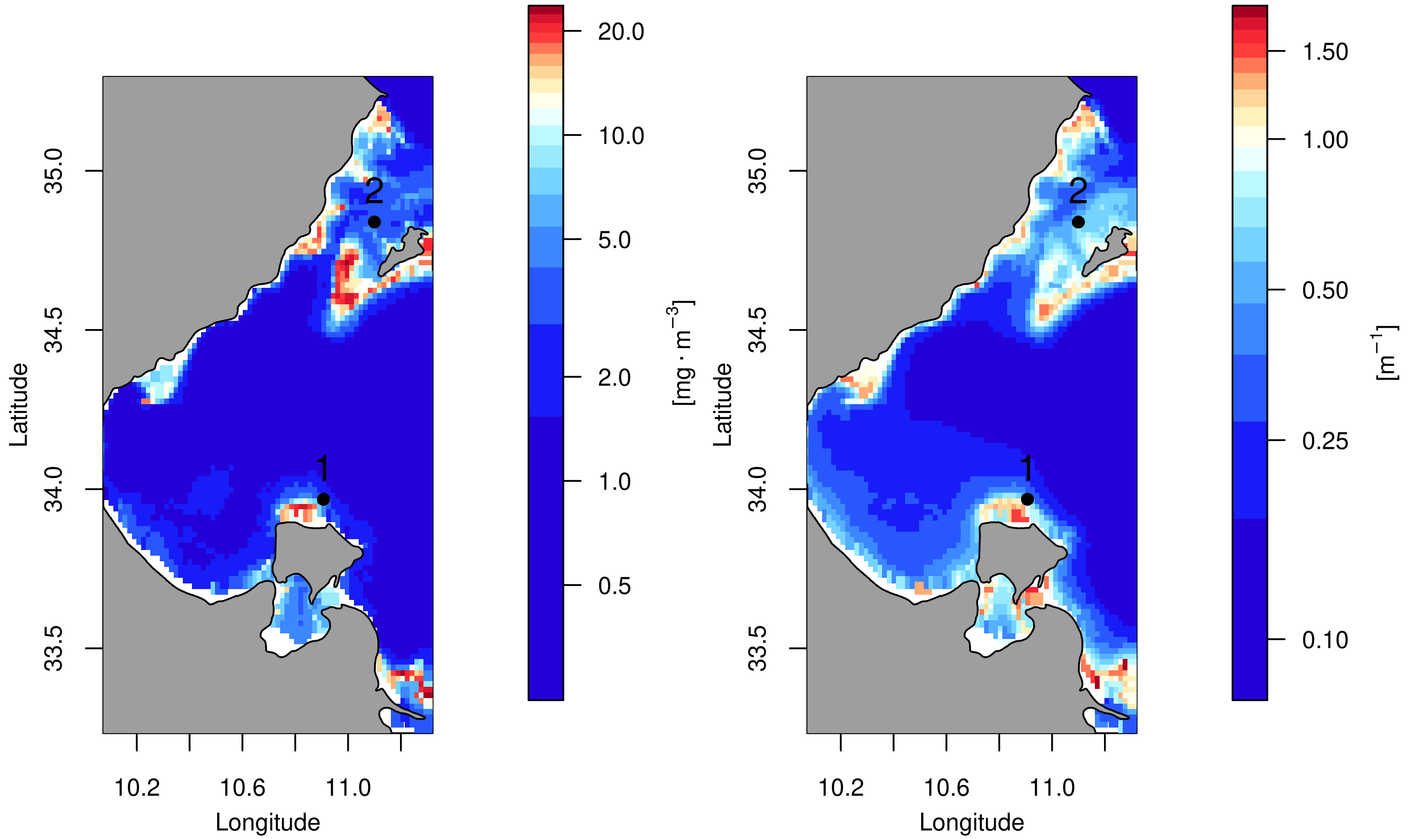}
	\end{tabular}
	\caption{Observed grid-points (4033 sites) in the Gulf of Gabes for the average Chl-a (left) and KD-490 (right) and two example sites reported in Figure \ref{fig:2}. \textcolor{black}{White color near the coastal area corresponds to sites with no values due to shallow waters.}}\label{fig:1}
\end{figure}

In Figure \ref{fig:1} we report the average levels of Chl-a concentration and KD-490 index for all the sites \textcolor{black}{over the spatial domain}.

\begin{figure}
	\centering
	\begin{tabular}{c}
\includegraphics[width=0.95\textwidth]{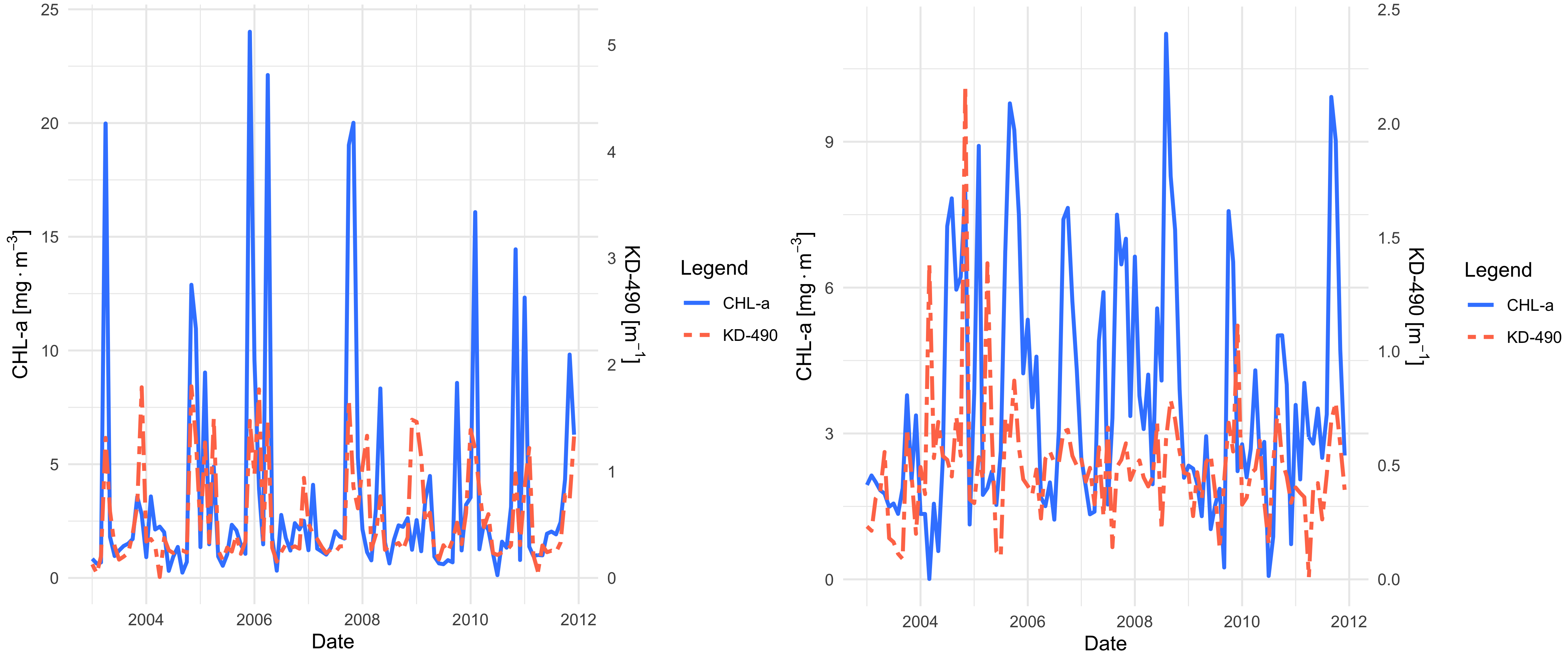}\\
\hspace{-1cm}(1) Lat: 10.88; Long: 33.99 \hspace{2.1cm} (2) Lat: 11.08; Long: 34.86\\
	\end{tabular}
	\caption{Temporal trend of Chl-a and Kd-490 in the two selected example sites.}\label{fig:2}
\end{figure}

The highest average levels of Chl-a and KD-490 were reported near the coastal area, in the north-eastern zone (Sfax) and in the southern part (Djerba) of the Gulf. The spatial distribution of the average values between Chl-a and KD-490 appears to have a similar spatial and temporal behaviour, but not everywhere. In fact as shown in Figure \ref{fig:2} the example site 1 exhibits a strong seasonal pattern both in the Chl-a and KD-490 indicators, with a strong correlation between the two time series (Pearson $\rho$: 66.1\%); in the second example site 2 there is a clear difference between the trend of the Chl-a concentration with respect to the KD-490 one as attested by a weak correlation (Pearson $\rho$: 30.7\%).

The dataset is affected by outliers due to measurement errors  (Chl-a concentration lower than 0.05 $mg/m^{-3}$: 1.0\%; KD-490 levels lower than 0.05 $m^{-1}$: 2.8\%) and by a  different amount of missing observations
(Chl-a: 19.1\%; KD-490: 1.8\%). 
The presented dataset presents several features which makes the clustering task challenging: a strong seasonality and a high variability,  the presence of outliers and a non bell-shaped distribution. The proposed classification may help to assess the trophic status of this area combining the information provided by different temporal Chl-a and KD-490 levels.
\section{Statistical modelling and inference}\label{sec3}
 \subsection{Multivariate quantile regression and asymmetric Laplace distribution}
We start by considering the  univariate case.
 Let $ Q_{p}(y|x) $, for $ 0<p<1 $, be the $p$-th quantile regression function of the univariate continuous random variable $ y $ given $ x$, a vector of  covariates.
 We suppose that $ Q_{p}(y|x)=x'\beta $, where $ \beta $ is a vector of unknown parameters to be estimated. Then a quantile regression model can be defined as $y=x'\beta+e$
where $ e$ is an error term  with  density function $f(\cdot;p) $ and  the $ p $-th quantile equal to zero, i.e. $ \int_{-\infty}^{0}f(e;p)de=p$.
Owing to the data $(y_t,x_t')$, $t=1,\ldots, T$, the  estimate of $\beta$ is classically \citep{koenker1978regression} obtained by minimizing 
\begin{eqnarray}\label{eq2}
	\hat{\beta}=\underset{\beta}{\text{argmin}}\sum_{t=1}^{T}\rho_{p}(y_{t}-x_{t}'\beta),
\end{eqnarray}
where $ \rho_{p}(\cdot) $ is the check loss function, i.e. $ \rho_{p}(z)=z(p-I(z<0)) $.

\citet{koenker1999goodness} showed that there is a direct relationship between minimizing (\ref{eq2}) and the maximum likelihood theory using independent variables $y_t$ with  Asymmetric Laplace   (AL) density
\begin{equation}
	f(e;p)=\frac{p(1-p)}{\sigma}\exp\left\{-\rho_{p}\left(\frac{e}{\sigma}\right)\right\},\label{eq3}
\end{equation}
where $ \sigma>0 $  is an additional   scale parameter.

In the following we will exploit the representation of $y$ with density \eqref{eq3} as a location scale mixture of Gaussian random variable  \citep{Kotz2001}, namely
\begin{eqnarray}
	y=x'\beta+\theta \sigma w+\omega \sigma\sqrt{ w}\,\nu \label{eq5}
\end{eqnarray}
where $ \nu\sim \mathcal{N}(0,1) $, and $ w $ is a exponential random variable with  $E(w)=1$. Here $\nu $ and $ w $ are mutually independent and
$\theta=(1-2p)/\{p(1-p)\}$ and $\omega^{2}=2/\{p(1-p)\}$.

The literature has focused on univariate response variable with only a few studies considering extension to multivariate case \citep{Benoit2012,Benoit2013,Waldmann2015,Petrella2019a}.

In particular \cite{Petrella2019a} considered a multivariate  asymmetric Laplace distribution \citep{Kotz2001} to specify a quantile regression model for the random vector
$\tilde{y}=(y_1,\ldots,y_q)'$ in the mixture representation
\begin{equation}
	\tilde{y}=X'\tilde{\beta}+ D\Theta {w} +\sqrt{w}D\Sigma^{1/2}\tilde{\nu} \label{eq3.20}
\end{equation}

Here  $\tilde{\nu}$ denotes a $q$-dimension standard Gaussian vector and $w$ is a exponential random variable with unit mean.
The matrix $X$ is a $q\times L$  regressor matrix and $\tilde\beta$ is a $L$-dimensional unknown vector. The other parameters of the model are contained in  $D = \operatorname{diag}(\tilde\sigma)$, with $\tilde{\sigma}=(\sigma_1,\ldots,\sigma_q)'$,  $\sigma_j>0$, $j=1,\ldots,q$  and $\Theta= \operatorname{diag}(\theta_1,\ldots,
\theta_q)'$ with $\theta_j = (1 - 2p_j)/[p_j (1 - p_j]$. 

Moreover the matrix $\Sigma=\Omega R(\tilde\phi)\Omega$ is a $q \times q$ positive definite matrix   with $R(\tilde\phi)$ being a correlation matrix, that depends on a vector of parameters $\tilde\phi$, and $\Omega= \operatorname{diag}(\omega_1^2,\ldots,\omega_q^2)$,
with entry $\omega_{j}^{2}={2}/[p_{j}(1-p_{j})]$. Note that using this formulation the component $y_j$ will  always be mutually dependent, even though $R(\phi)$ is an identity matrix.

In this paper we  consider a slight different approach with respect to  \eqref{eq3.20}.
We use the same idea as in \cite{Waldmann2015} and we set

\begin{equation}\label{eq:ald}
	\tilde{y}=X'\beta+ D\Theta \tilde{w} +{W}^{1/2}D\Sigma^{1/2}\tilde{\nu} 
\end{equation}
with $\tilde{w}= ({w}_1,\ldots,w_q)'$ and ${W}= \operatorname{diag}({w}_1,\ldots,{w}_q)$
allowing a different  exponential random variable with unit mean $ w_j$ for each component.

With this variation   model \eqref{eq:ald} does not define a proper multivariate asymmetric Laplace distribution as in \citet{Kotz2001}. However the model becomes more flexible and can cover the case of mutually independent components.
Now the challenge  lies in finding a multivariate density
for $\tilde{w}$, $	e(\tilde{w};\gamma)$, with marginal exponential distributions for the $w_j$ in order to guarantee that the marginal distributions for the response variables are still asymmetric Laplace. Here the parameter $\gamma$ is a generic dependence parameter.

Several multivariate exponential distributions   have been proposed in the literature \citep[Ch. 47]{kotz:balakrishnan:johnson}.
\cite{Waldmann2015} adopt the simplest specification that the exponential variables are independent, i.e. $\gamma=0$.

In view of the bivariate real data example in this paper we exemplify our construction by using the bivariate density proposed by \cite{Downton1970}, namely
\begin{equation}\label{eq:downton}
	e(\tilde{w};\gamma)= \frac{1}{(1-\gamma)}\exp \left\{-\frac{1}{1-\gamma}\left(w_{1}+w_{2}\right)\right\}
I_{0}\left(\frac{2}{1-\gamma}\sqrt{\gamma w_{1}w_{2}}\right)
\end{equation}
where $ 0\le\gamma < 1 $
and $I_{0}(a)=\sum_{k=1}^{\infty}\frac{a^{2k}}{4^{k}(k!)^{2}}$ is the modified Bessel function of the first kind of order zero.
The value $\gamma$ represents the 
Pearson's product-moment correlation and $ \gamma=0 $ implies independence between the $w_1$ and $w_2$.
Coupling \eqref{eq:ald} with the density \eqref{eq:downton} we obtain a  flexible specification of a bivariate quantile regression model that covers the case of independence between  the components when $R(\phi)$ is an identity matrix and $\gamma=0$.

\subsection{Model-based clustering}\label{subsec:mixmod}
Suppose that we observe  data over $n$ statistical units and, for simplicity,  the same number $q\times T$  of values $y_{ijt}$,  $j=1,\ldots, q$, $t=1,\ldots,T$, for each statistical unit $i$, $i=1,\ldots,n$.
We collect the observations for each unit into the vector $\boldsymbol{y}_i = (\tilde{y}_{i1}',\ldots, \tilde{y}_{iT}')'$, with
$\tilde{y}_{it} = (y_{i1t},y_{i2t},\ldots y_{iqt})'$, $t=1,\ldots,T$.
The vectors  $\boldsymbol{y}_i$ are supposed to be independent.
The whole dataset will be denoted by $\mathcal{Y}=(\boldsymbol{y}_1,\ldots,\boldsymbol{y}_n)$.

Our statistical  problem is to cluster the $n$ statistical units, i.e. the sites,  into the clusters $\{1,\ldots,K\}$, $K\ll n$.
We follow a  mixture-model  approach \citep{mclachlan:peel:2000}  to clustering  according to which the cluster membership of the $i$th unit is represented by a latent random variable  ${c}_i \in \{1,\ldots,K\} $ , where $c_{i} = k$ indicates if the $i$-th units belongs to the cluster $k$. 

The memberships ${c}_i$, $i=1,\ldots,n$ are supposed independent and identically distributed  variables with 
$\Pr(c_{i} = k)=\alpha_k$, $ 0<\alpha_{k}<1 $, for all $k=1,\ldots,K$ and $ \sum_{k=1}^{K}\alpha_{k}=1 $.

Given the cluster membership,  $c_{i}$,  observations for the $i$-th unit are generated by the mixture model specified hierarchically, namely 

\noindent\textbf{the observation level}
\begin{equation}\label{eq:model}
		\tilde{y}_{it}|c_i,\tilde{w}_{it}\sim\mathcal{N}_q(X_{t}'\tilde\beta_{c_i}+ D_{c_i}\Theta \tilde{w}_{it},{W}_{it}D_{c_i}\Sigma_{c_i}D_{c_i}{W}_{it}), \qquad i=1,\ldots,n,\quad t=1,\ldots,T 
\end{equation}
		conditionally independent distributed, with $\Sigma_{c_i}=\Omega R(\phi_{c_i})\Omega$;

\noindent \textbf{the latent process level}
\begin{equation}
		\tilde{w}_{it}|c_i\sim e(w,\gamma_{c_i}), \quad i=1,\ldots,n,\quad t=1,\ldots,T
\end{equation}
		conditionally independent distributed
\begin{equation}
		c_i \sim \operatorname{Multinomial}(1,{\alpha}), \quad i=1,\ldots,n
\end{equation}
independent distributed.  
\textcolor{black}{However, in presence of statistical units over a spatial domain, the incorporation of the spatial dependence in the model can be a feature that needs to be considered. Since the aim of our main application (see Section \ref{sec5}) was to discover sites at different environmental pressure, the incorporation of spatial dependence in the MCMC chain can eventually mask the classification by a common smoothing effect. Depending on the type of spatial domain and the analysed data,  this dependence can be incorporated, for example, by considering a Markov random field as, for example, in \citep{Gaetan:2017,jiang2012clustering} that takes into account the membership of the nearest neighbors.}

\subsection{Bayesian inference}\label{subsec:bayinf}
We adopt a Bayesian approach to make inference on the model parameters.
The inference is facilitated by the fact that the mixture model is  specified hierarchically. Moreover
the conjugacy of some priors leads to  updates with
simple and established methods for drawing from the full
conditional distribution.

\noindent \textbf{Prior distributions}

A common choice for the prior distribution of $\alpha$ is $ {\alpha}\sim \mathcal{D}(a_{1},...,a_{K}) $, where $\mathcal{D}$ is the Dirichlet distribution with parameters $a_1, \ldots, a_K > 0$;  

The choice $ \tilde\beta_{k} \sim \mathcal{N}(b_k,P_k),\, k=1,\ldots,K$, identically and independent distributed is still common and simplifies the simulation. Setting $ b_k=0 $ and $ P_k=a\,I $, for $ a \gg 0 $, leads to an improper prior.
		
		 Specification of the  prior distributions for
		$\tilde\sigma_1,\ldots,\tilde\sigma_k$ and $\tilde\phi_1,\ldots,\tilde\phi_k$  is complicated  by the complex requirement that the matrices $D_k\Omega R(\phi_{k})\Omega D_k$ are  non-negative definite matrices.
We follow \citet{barnard:mcculloch:meng:2000} and   work  by specifying prior for variances  and correlation matrices.
For the variances we put the priors $ \sigma^2_{kj} \sim \mathrm{IG}(s_{k},d_{k}),\quad k=1,\ldots,K,\quad j=1,\ldots,q$,  identically and independent distributed,
	where $IG$ is  the inverse gamma distribution,
	with the shape $ s_{k}$  and $ d_{k}$ scale parameters.
	\citet{barnard:mcculloch:meng:2000} discussed the relative merits of choosing a prior for $\tilde\phi_1,\ldots,\tilde\phi_k$ independent from $\tilde\sigma_1,\ldots,\tilde\sigma_k$.
 In our example the choice is greatly simplified since we consider two variables, i.e. $q=2$. In that case
we can assign assign a uniform prior, $\phi_k\sim\mathcal{U}(-1,1)$,  for the correlation coefficient $\phi_k$ that are supposed identically and independent distributed.

For sampling
from the posterior distribution we use a hybrid MCMC algorithm known as Metropolis-within-Gibbs algorithm \citep[][Chapter 10]{Robert2004}.
As we show in the Supplementary material the conjugacy of some priors leads to  updates with
simple, established methods for drawing from the full
conditional distribution.
In other cases, we  resort to Metropolis-Hastings
to draw from some of the full conditional distributions.

\section{A simulation study}\label{sec4}
\noindent

In this simulation 
study 
we want to exemplify how the mixture model described in subsection \eqref{subsec:bayinf} is able to cluster different bivariate temporal patterns encountered.

We will consider two experiments. A first experiment (Sim A) in which data are generated   from  a symmetric  distributions and a second example (Sim B) in which data come from asymmetric distributions.
In both experiments we assume that $n=300$ statistical units are split into three clusters of size $n_k=100$, $k=1,2,3$.
For each unit we simulate a bivariate vector of length $T=100$
$(y_{1t},y_{2t})'$, $t=1,\ldots,T$ with  time-varying marginal distributions.

Let $z_{jt}$, $j=1,2$, $t=1,\ldots,T$ a standardized Gaussian random variable and $m_{jk}(t) $ a positive function, $k=1,2,3$, $j=1,2$, $t=1,\ldots,T$.
We consider two setting for simulating $(y_{1t},y_{2t})'$, namely

\begin{enumerate}
	\item[Sim A:]  ${y}_{jt}=m_{jk}(t)+\sqrt{m_{jk}(t)/5}\, z_{jt}$;
	\item[Sim B:] ${y}_{jt} = G^{-1}(\Phi(z_{jt});m_{jk}(t)/5,5)$, where
	$\Phi(z)$ is the cumulative distribution function (CDF) of a standardized Gaussian random variable and $G^{-1}(u;a,b)$ is the inverse of the CDF of a Gamma random variable
	with mean $ab$ and variance $ab^2$.	
\end{enumerate}
The $k$ value  in $m_{jk}(t)$ varies between 1 and 3 depending on the cluster membership.
Note that  the marginal distributions were chosen in the way that the means and the marginal variances are equal in the two settings.

The  temporal patterns in each cluster are led by  the function
$
g(t;a,b)=a[2+t/T + \text{exp}\{-(t/T-b)^{2}/0.05\}]
$, that is
\begin{center}
	\begin{tabular}{ccc}
\textbf{Cluster}		&\multicolumn{2}{c}{\textbf{Component}} \\
\hline
		& $j=1$ & $j=2$ \\
		\hline
		$k=1$ & $m_{11}(t) = g(t;1,0.2)$ &$ m_{21}(t) = g(t;1.5,0.8)$\\
		$k=2$ & $m_{12}(t)=g(t;1,0.5)$&$ m_{22}(t)=g(t;1.5,0.2)$\\
		$k=3$& $m_{13}(t)=g(t;1,0.8)$& $ m_{23}(t)=g(t;1.5,0.5)$\\
		\hline
	\end{tabular}
\end{center}

In order to asses the robustness of the procedure in the presence of serial dependence,  we simulate the bivariate time series $(z_{1t},z_{2t})'$ as
 $z_{jt}=v_{jt}+\theta v_{jt-1}$,  $j=1,2$,
where 
$(v_{1t},v_{2t})'\sim\mathcal{N}(0,\Sigma_v)$,  is a bivariate white noise, with $\displaystyle \Sigma_v=\frac{1}{1+\theta^2}\left[\begin{array}{cc}
	1& \rho\\
	\rho & 1
\end{array}\right]
$,
  $-1<\rho<1$ and $-1\le\theta\le 1$. By choosing different  values for  $\rho$ and $\theta$, different degrees of mutual and serial dependence are obtained.

In order to capture the temporal component of the  bivariate vector in each cluster, we follow a regression spline approach.  We choose a cubic B-spline basis, $b_1(t),\ldots,b_{m}(t)$, with equally spaced knots over the range of time. For simplicity,
 we assume the same number $m$ of basis functions for both component of $\tilde{y}_t$.  The resulting matrix $X_{t}$ of regressors  in \eqref{eq:model} 
is
$\displaystyle
X_{t}=\left(
\begin{array}{cc}
x_{t}& 0\\
0 & x_{t}\\
\end{array}
\right)'
$
with $x_t=(b_1(t),\ldots,b_{m}(t))'$.

We suggest this simple strategy for a  preliminary selection of the number of basis $m$, namely
\begin{enumerate}
\item fix the value $p_j$, $j=1,2$ in \eqref{eq2}	and get $\hat{\beta}_j$, $j=1,2$, for each time series $\{y_{jt},\,t=1,\ldots,T\}$;
\item evaluate the AIC-like criterion
$
AIC_j(m)=\sum_{t=1}^{T}\rho_{p}(y_{jt}-x_{t}'\hat\beta_j)+2 m
$;
\item repeat step 1 and 2 for each   statistical unit $i$ and obtain the value $AIC_j^{(i)}(m)$, $i=1,\ldots,n$;
\item find the value $m$ that minimizes the overall  value
$\overline{AIC}(m)=\sum_{i=1}^n\sum_{j=1}^2 AIC_j^{(i)}(m)$.
\end{enumerate}	
In our simulation study  we consider the median value for both  time series, i.e. $p_1=p_2=0.5$.

For estimating the model parameters $\boldsymbol{\psi}$, we run  the MCMC for $100$ iterations as burn-in and $300$ iterations for getting the posterior estimates. Inspection of trace plots suggests convergence of the parameters. 
 From the clustering partitions sampled in the MCMC algorithm we obtain an estimate of the clustering structure by considering the posterior mode.

Both experiments are repeated  $100$ times.  We compared the performance of our clustering method with three state-of art competitors:
\begin{itemize}
	\item [1)] Gaussian finite Mixture model Clustering (GMC): we cluster the data by means of a mixture of Gaussian linear regression models. For estimating the parameter we exploit the R package \texttt{flexmix} \citep{flexmix} which is based on Expectation-Maximization  algorithm;
	\item [2)] Raw Data Clustering (RDC): each bivariate time series is stacked in one vector. Then the vectors are clustered by means of a Partitioning Around Medoids (PAM) algorithm, extracting three clusters;
	\item[3)] CHaracteristic-based Clustering (CHC). A global measure describing the time series is obtained by applying summary indices about trend, seasonality, periodicity, serial correlation, skewness, kurtosis, chaos, nonlinearity and self-similarity \citep{Wang2006}. The normalized indices or features extracted by  using the R package \texttt{tsfeatures} \citep{tsfeatures}
	 are the inputs of the PAM algorithm. 
	
\end{itemize}

We assess the power of our clustering algorithm in reconstructing the three clusters by comparing the level of agreement between the estimated partition and the true clustering using 
the Adjusted Rand Index, ARI, \citep{Hubert1985}.

 Table \ref{bivsimARI} presents the ARI values for the combinations of the correlation structures and the marginal distribution.
From these results we note that our bivariate clustering algorithm performs very well at the combination, reporting low clustering performance for the more complex structure in particular for $\phi$=0.5; the simulation with the Gamma distribution (Sim B) appears more challenging in the clustering for all the considered methods.  It is worth to note that our method reported a better classification than the Gaussian mixed linear regression model even in the case of normal marginal distribution, especially in presence of a serial dependence.

\begin{table}[h!]\centering\vspace{0.5cm}
	\arrayrulecolor{black}
	\medskip
	\begin{tabular}{lcccccc}
		\hline
			 &$\rho$&$\theta$& {Our method} & GMC & RDC & CHC  \\
		\hline
		Sim A	&	$0$&$0 $ & $ 1.00\, (0.01) $ & $ 0.94\, (0.10) $& $ 1.00\, (0.01) $& $ 0.82\, (0.08) $\\
		&	$ 0.5$&$0  $ & $ 1.00\, (0.01) $ & $ 0.95\, (0.09) $& $ 1.00\, (0.01) $& $ 0.80\, (0.05) $\\
		
		&	$ 0.5$&$1.0  $ & $ 0.91\, (0.15) $ & $ 0.57\, (0.06) $& $ 0.75\, (0.09) $& $ 0.57\, (0.07) $\\
		\hline
		Sim B	&	$ 0$&$0 $ & $ 0.84\,(0.12) $ & $ 0.44\, (0.05)  $& $ 0.35\, (0.11) $& $ 0.55\, (0.09) $\\
		&	$ 0.5$&$0  $ & $ 0.56\, (0.29)  $ & $ 0.44\, (0.05) $& $ 0.35\, (0.10) $& $ 0.58\, (0.06) $\\	
		&	$ 0.5$&$1.0  $ & $ 0.32\, (0.22) $ & $ 0.32\, (0.05) $& $ 0.22\, (0.10)  $& $ 0.42\, (0.06) $\\
		\hline
	\end{tabular}
	\caption{Average (standard error between parentheses) Adjusted Rand Index values of 100 replications for each $\rho$ and $\theta$  combination, marginal distribution and clustering method.\label{bivsimARI}}

\end{table}

\section{Clustering sites in the Gulf of Gabes}\label{sec5}

In this section we present the clustering results for the bivariate variables Chl-a concentration and the KD-490 levels, previously presented in Section \ref{sec2}.  
As reported in Figure \ref{fig:2} and \ref{fig:2plus}, we can observe a strong and time-varying seasonal pattern, with the presence of a peak at the beginning of each year. For this reason
 we choose to model such monthly seasonality using sine and cosine functions \citep{eilers2008modulation}.
 More precisely we suppose that for each variable $y$ the quantile regression function is a function of the time $t$, such that 
 \begin{equation*}
Q_p(y|t)=g_1(t)+ g_2(t) \cos(\pi/6t)+ g_3(t) \sin(\pi/6t)
 \end{equation*}
 where   a possible overall trend is represented by a smooth function $g_1(t)$, while $g_2(t)$ and $g_3(t)$ are smooth functions  that modulate the local amplitudes of the cosine and sine waves. 

In order to have a good grade of flexibility, the three functions $g_1(t)$, $g_2(t)$, and $g_3(t)$ take the form of a regression on a cubic B-spline basis, $
g_j(t)= \sum_{l=1}^{m_j} \beta_{jl} b_l(t)
$
with  equally spaced knots over the time interval $[1,108]$.
It is easy to see that the resulting  model for the quantile function
 \begin{equation*}
	Q_p(y|t)=\sum_{l=1}^{m_1} \beta_{1l} b_l(t)+ \sum_{l=1}^{m_2} \beta_{2l}[b_l(t)\cos(\pi/6t)]+ \sum_{l=1}^{m_3} \beta_{3l}[b_l(t)\sin(\pi/6t)]
\end{equation*}
can be written as a linear combination of covariates  that fits with \eqref{eq:ald}.

We carry out a preliminary data analysis in order to get the degree of smoothing and the number of clusters.
This data  analysis has been performed on the time series  of Chl-a concentration since this variable displayed more heterogeneity in space and time.

To find the  grade of smoothing and obtain a value for $m_1$, $m_2$, and $m_3$, we follow the strategy outlined in Section \ref{sec4} and we minimize the overall value $AIC(m)$, with  $m=m_1+m_2+m_3$  by means of a median (i.e. $p=0.5$) regression \textcolor{black}{and considering a number of basis for each component from 3 to 6.}
The solution with $m_1=4$ for the  trend (with the inclusion of a internal intercept) and $m_2=m_3=3$ for the cyclical components minimized the overall $AIC(m)$ and it was chosen in the following models.

\textcolor{black}{For a fixed number  of clusters $K$, we fit a bivariate model assuming that the two time series have a constant correlation coefficient across the clusters, i.e. $\phi=\phi_k$.
Moreover a word of caution is in order of the estimate of the parameter $\gamma$ in \eqref{eq:downton}. Our experience with this dataset indicates that the data provide very little information on the parameter. For this reason we set the value $\gamma=0.5$, giving the parameter $\phi$ the task of modulating the dependence between the two time series.
}

\textcolor{black}{The results were obtained after 2500 Monte Carlo iterations using a burn-in of 300. The final membership ${c}_i$ and the regression coefficients $\beta_{jl}$ were estimated by means of the mode and mean of a-posteriori distribution, respectively.}

\textcolor{black}{We identify $K$ minimizing an adapted  version of the Deviance Information Criterion ($DIC$) \citep{spiegelhalter2002bayesian} following \citet{celeux2006deviance}. More precisely starting from the formulation   named $DIC_2$ in that paper
}
$$
DIC_{2}(\mathbf{y},\mathbf{\psi})= -4\mathbb{E}_{\psi,\mathcal{C}}[\text{log}f(\mathbf{y}|,\psi)|\mathbf{y}]+2\text{log}f(\mathbf{y},\tilde{\psi}(\mathbf{y})) \label{eq2.27}
$$
\textcolor{black}{
where $\mathbf{\psi} =(\mathbb{\beta},\sigma)'$ and $\tilde{\psi}(\mathbf{y})$ is the  maximum a posteriori (MAP) estimates of ${\psi}$.   \citet{celeux2006deviance}  approximates $ DIC_{2} $  by using the MCMC runs 
}

\begin{eqnarray}
CDIC(\mathbf{y})&=& -\frac{4}{m}\sum_{l=1}^{m}\sum_{i=1}^{n}\log\left\{\sum_{k=1}^{K}
\alpha_{k}^{(l)}
 f(\mathbf{y}_{i}|\psi_k^{(l)})\right\}\nonumber\\
&& +\quad 2\sum_{i=1}^{n}\log\left\{\frac{1}{m}\sum_{l=1}^m \sum_{k=1}^{K}\alpha_{k}^{(l)} f(\mathbf{y}_{i}|\psi_k^{(l)})\right\}\label{eq:dic}
\end{eqnarray}
\textcolor{black}{where $ \psi_k^{(m)} $ and $ \alpha_{k}^{(l)}$ are the results of the $l$-th MCMC iteration.}

\textcolor{black}{However, Formula \ref{eq:dic} entails the evaluation of  the bivariate density function 
$f(y_{i1t}, y_{i2t};\psi_k)$ that arises from equation \eqref{eq:ald} by integrating out the random variable $\tilde{w}$. Since the bivariate  density function cannot be derived in closed form, we proposed a composite version of the DIC index, called Composite-DIC (CDIC), pretending in \eqref{eq:dic} that  $f(y_{i1t}, y_{i2t};\psi_k)=f( y_{i1t};\psi_k)\cdot f(y_{i2t};\psi_k)$ in the same spirit of \cite{varin2005}.}

 We estimate the model for a range of different quantile combinations namely for different values of $(p_{1},p_{2})$. In particular we consider the combination of quantiles given by the pairs $ (0.5,0.5) $, $ (0.9,0.5) $, and $ (0.9,0.9) $ for Chl-a concentration and KD-490, respectively. \textcolor{black}{While the quantile 0.5 can represent a robust estimate of the central tendency of the behaviour of each indicator, the quantile 0.9 is particularly important in ecology to evaluate the temporal trend towards the upper end of the distribution.}

\textcolor{black}{The values of  CDIC for a number of clusters $K$ which varies from 2 to 7 are reported in Table \ref{tab_cdic}. }  

\begin{table}[ht]
\centering
\textcolor{black}{
\begin{tabular}{cccc}
 \hline
 $K$ &    {(0.5,0.5) }& {(0.9,0.5) }&    {(0.9,0.9)}\\
 \hline
7 & 33.1   & 46.1     & 48.4 \\
6 & \textbf{31.2}        & 44.1      & 46.5      \\
5 & 33.1     & 44.0      & 48.2   \\
4 & 35.1       & \textbf{39.8} & 46.9   \\
3 & 32.3  & 44.1     & \textbf{46.4}     \\
2 & 39.5   & 41.0        & 46.6\\       
 \hline
\end{tabular}
}
	\caption{\textcolor{black}{CDIC ($\times 10^5$) values for $K$  at different quantile pairs. The lowest CDIC value is highlighted in bold.} \label{tab_cdic} }
\end{table}
\textcolor{black}{Considering the values reported in Table \ref{tab_cdic} we have chosen a number of clusters equal to K = 6, 4 and 3 for the quantile combination $ (0.5,0.5) $, $ (0.9,0.5) $ and $ (0, 9,0,9) $, respectively. This result suggests  a different number of clusters for each combination indicating a decreasing number to an increasing combination of quantile levels. Especially for the latest model, the proposed criterion suggests a classification based on three clusters.}

\begin{figure}[ht]
  \centering
\includegraphics[width=0.95\linewidth]{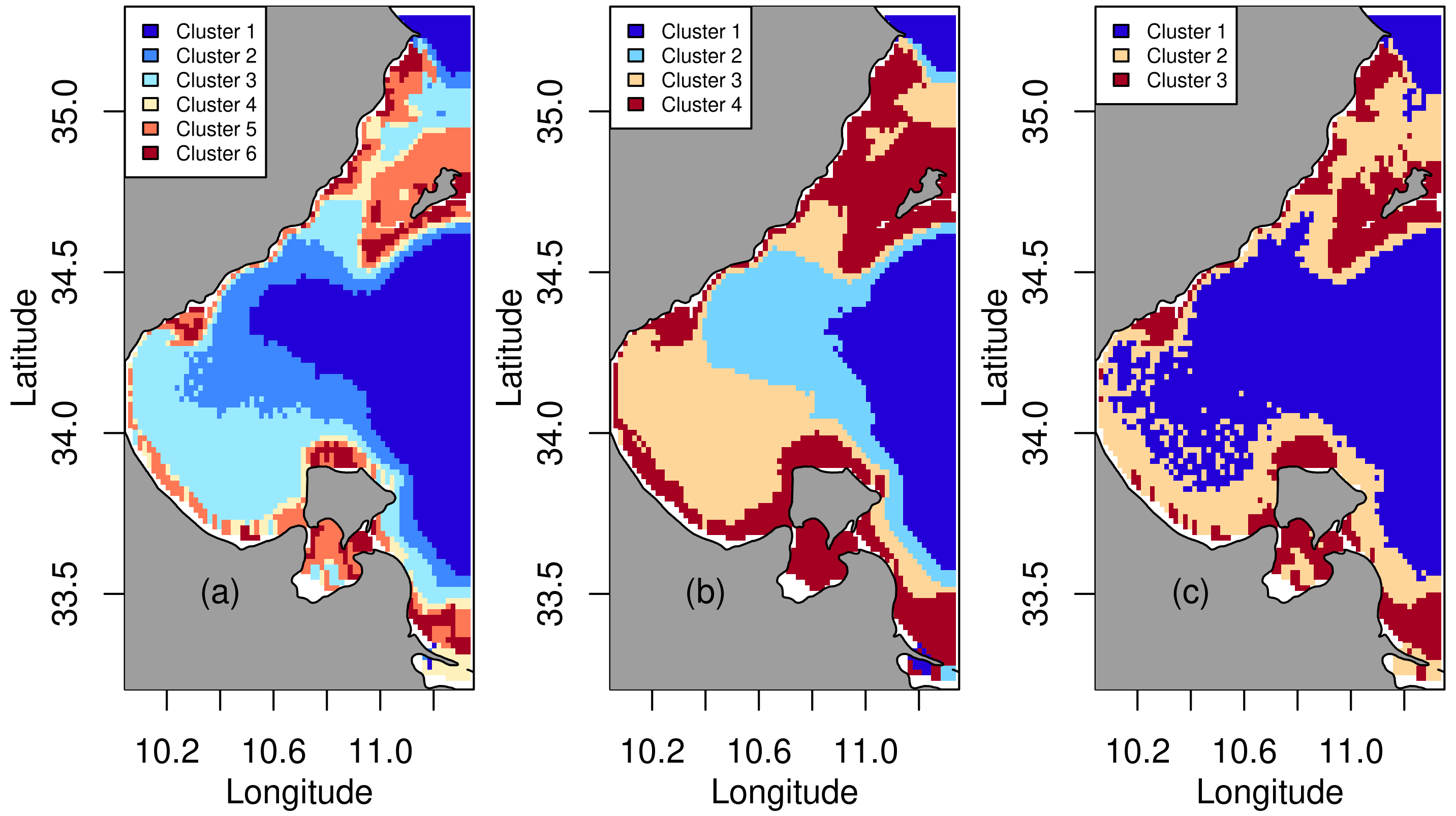}
	\caption{Spatial clustering results for Chl-a concentration and KD-490 index for each quantile combination $(p_1,p_2)$: (a)  $ (0.5,0.5)$ (left), (b) $ (0.9,0.5)$ (middle), and (c) $ (0.9,0.9)$ (right). \label{spatial} }
\end{figure}

In Figure \ref{spatial} we present the spatial distribution of the clustering results for each combination of quantiles.

Across all the fitted quantile combinations, we observe that those regions around the islands (Kerkennah, Kneiss and Jerba) are clustered as having the highest average values of both Chl-a and KD-490 concentration which decreases as one moves towards the deep sea. These results are consistent with the results obtained from the univariate case as well as the findings of \cite{Katlane2012} who notes that, from multi-temporal turbidity maps produced from Moderate Resolution Imaging Spectrometer (MODIS) for $ 2009 $, areas around the islands (Kerkenah, Kneiss and Jerba) and at the industrial port of Gannouch, were characterized by high turbidity variation, concentration of total suspended matter and Chl-a concentration.

\begin{figure}[!ht]
	\centering
	\includegraphics[width=1\textwidth]{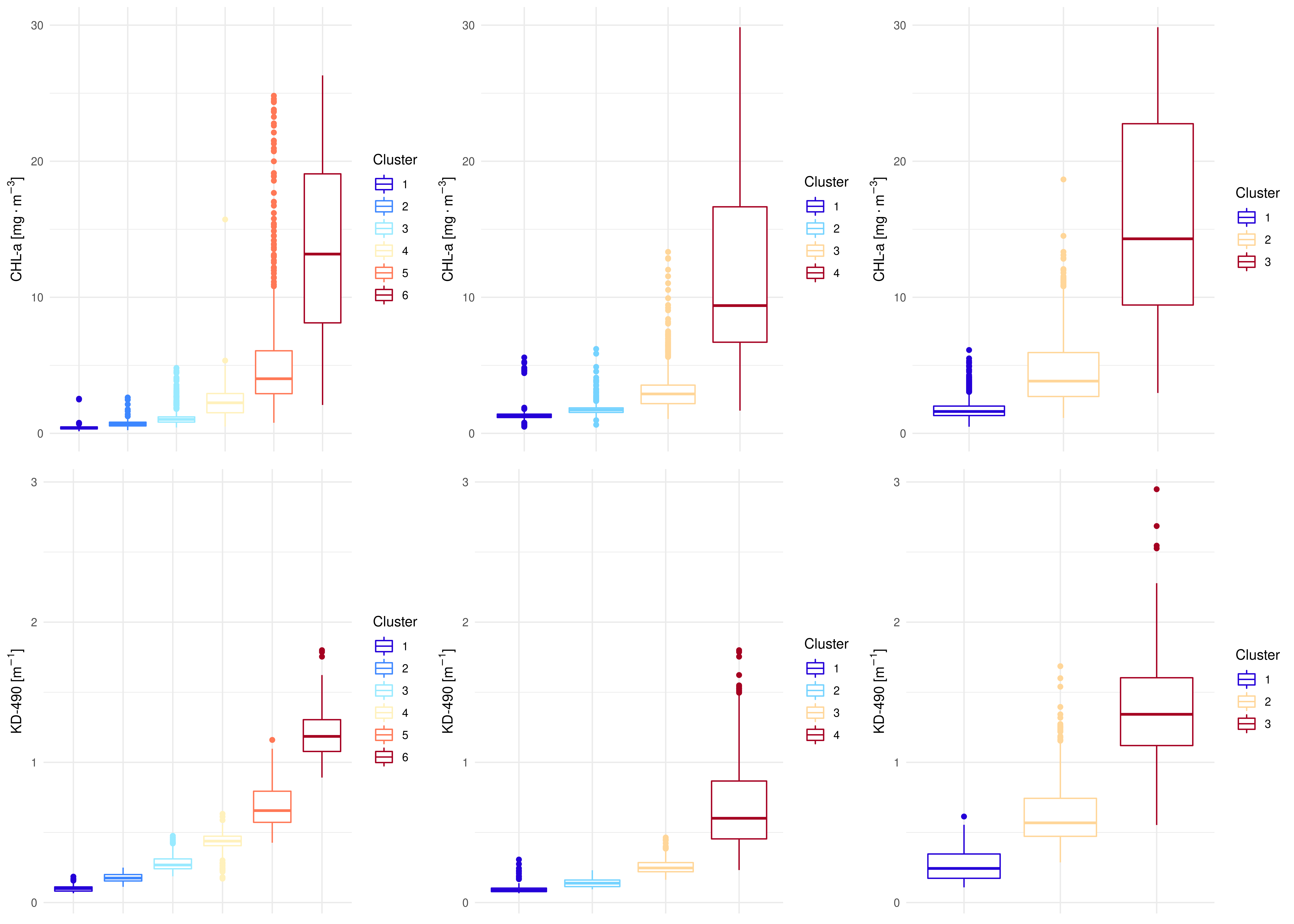}\\
	\caption{Distribution of \textcolor{black}{reference quantile} of Chl-a concentration and KD-490 index by Cluster for each quantile combination $(p_1,p_2)$:  $ (0.5,0.5)$ (left), $ (0.9,0.5)$ (middle) and $ (0.9,0.9)$ (right). \label{boxplots} }
\end{figure}

\textcolor{black}{However, given the different meanings of Chl-a concentration and KD-490 levels at each quantile level, the spatial distribution between the three classifications appears different in particular looking at the northeastern area and the coastal zone. The classification with the quantile combination ($0.5,0.5$) identifies a series of clusters that grades not only the most polluted area (Cluster $6$) but also the coastal area (Cluster $4$ and $5$). The zone inside the Gulf is classified with the Cluster $3$, while the remaining zone far from the coastal area and islands is covered by the Cluster $1$ and $2$. Otherwise, the classification with ($0.9,0.5$) reported a similar spatial extension concerning the Cluster $1$ and $2$ which include zones starting from the offshore waters to the coastal area, while the coastal zone and the Sfax industrial area in the north-east are entirely covered by the Cluster $4$; the Cluster $3$ defines a transition zone. As reported in Figure \ref{boxplots}, both the classification performed with ($0.5,0.5$) and ($0.9,0.5$) report an increasing trend in the observed values of both Chl-a concentrations and KD-490 levels with the increase of the cluster label. 
The clustering performed with a quantile combination of 0.9 for Chl-a and 0.5 for KD-490 appears different with respect to the combination $(0.5,0.5)$ both in terms of marginal and spatial distribution: in the quantile combination $(0.9,0.5)$ Cluster $4$ embraces an increased percentage of the seawater (24.6\%) and it shows lower values with respect to the previous most impacted area (Cluster $6$) obtained by the classification with quantiles $(0.5,0.5)$; low KD-490 and CHL-a concentrations are restricted for both the classifications ($(0.5,0.5)$ and $(0.9,0.5)$) to the Cluster $1$, $2$ and Cluster $3$. In the classification with $(0.9,0.9)$, the number of sites that belongs to Cluster $1$ increases to 56.3\%, and the relative spatial extension covers all the offshore water area. The Chl-a and  KD-490 values in the Cluster $1$ are low and widely separated from the Cluster $2$ in comparison with the previous classifications. The sites belonging to Cluster $2$ cover in a unique group all the sites previously classified as a transition zone or near the coastal area. The Cluster $3$ identifies four marine areas heavily impacted by high measurements of Chl-a and KD-490: the seawater around Jerba, Sfax, Zarzis, and Sharqi Island. }

\begin{figure}[htp]
	\centering
	\includegraphics[width=0.95\linewidth]{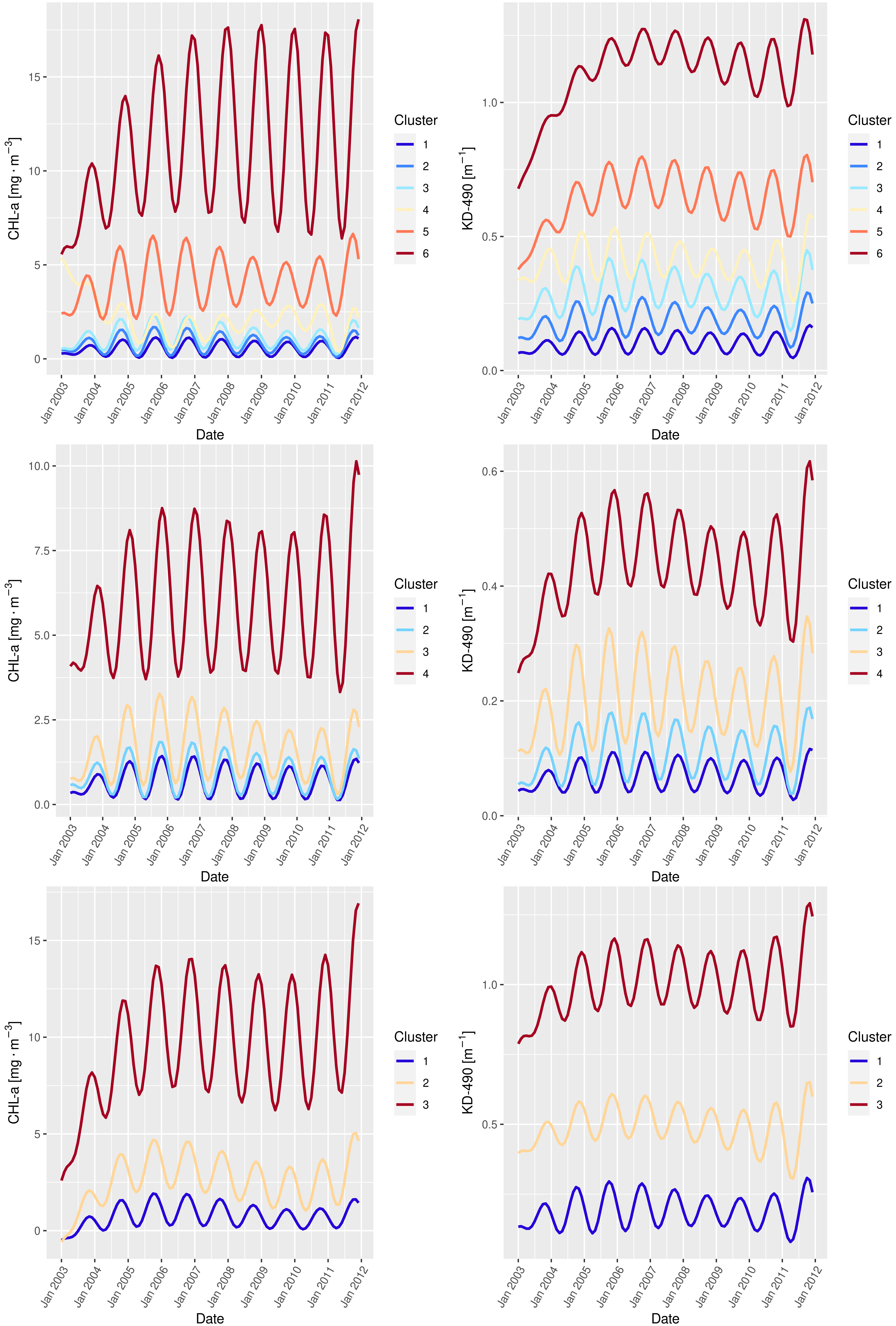}\\
	\caption{\textcolor{black}{Estimated temporal component  for Chl-a concentration and KD-490 index at quantile combinations $(p_1,p_2)$:  $ (0.5,0.5)  $ (top),  $ (0.9,0.5)$ (middle), and   $ (0.9,0.9) $ (bottom). \label{timeseries} }}
\end{figure}

\textcolor{black}{In Figure \ref{timeseries} the temporal pattern for each recovered cluster is reported by plotting the estimated temporal component ($\hat\beta X$) within each identified group of Chl-a concentration and KD-490 level for each considered combination of quantiles. For all clusters a seasonal pattern is evident. Cluster labels and colors are the same as those in Figure \ref{spatial} \textcolor{black}{ and they are ordered by increasing average Chl-a concentration.}
The classification with the quantile combination $(0.5,0.5)$ is guided by different average levels for Chl-a concentration and KD-490 level; however, especially for Chl-a concentration, the first three clusters ($1$, $2$, and $3$) look very close; the presence of a different classification for those groups is explained by a clear separation considering  respective KD-490 levels: posing our attention to KD-490, the curves report different average values, but a similar seasonal component. In addition, the Cluster $6$ reports an increasing trend at the beginning  of the temporal window for both the indicators and a strong cyclical pattern in Chl-a concentration, while the seasonality is less evident for KD-490.}

\textcolor{black}{In Figure \ref{boxplots}, considering the two boxplots in the middle related to the clustering results obtained  with the  quantile combination $(0.9-0.5)$, the analysis of the differences between the estimated groups reports as sites belonging to  Cluster $4$ are those with the highest values and a well separated from the other clusters for both the indicators. Otherwise, Cluster $1$, $2$, and $3$ are close together and the difference is mainly due to a different average trend, a greater intra-season amplitude for Cluster $3$, and a late seasonal peak for Cluster $1$, as denoted by the Figure \ref{timeseries}.
In addition, the temporal trend appears to be different from the previous classification: only for KD-490,  Cluster $4$ reports an evident increasing trend, followed by a stabilization and a slight decrease after the year 2008.
Taking into account the clustering results obtained by the quantiles $(0.9-0.9)$,  all clusters are well separated. Cluster $1$ reports the lowest values of limited signal amplitude, while the Cluster $2$ shows a stable and cyclical behaviour, far from other groups. Cluster $3$ exhibits an increasing trend and a stabilization after 2006 only, for Chl-a concentrations, while the KD-490 trend is stable and cyclical.}
\section{Discussion}\label{sec6}
In this paper we have proposed a new model-based clustering technique,  that is capable of handling asymmetric clusters with the presence of outliers as well as considering different quantile levels of the observed data. Our clustering strategy is based on the finite mixture model theory where each component of the mixture of AL distributions, which constitutes  the density of bivariate random variables which are potentially correlated, is assumed to represent a cluster with the skewness parameter of AL distribution being used to directly model the quantiles of interest. Therefore our proposed technique adds to the richness of the recent burgeoning of non-Gaussian approaches to model-based clustering.

\textcolor{black}{As the AL distribution provides a direct link between the maximum likelihood theory and minimization of a quantile regression check loss function \citep[see][]{koenker1999goodness,yu2001bayesian}, we estimate the cluster-specific  parameters, which are the parameters of the mixing AL distribution, through a Bayesian 
approach.} 

In our simulation experiment we considered three clusters two of which are not distinctively different from each other. However, an evaluation of the power of our proposed algorithm in reconstructing the three groups indicates a good performance with respect to the other competitive methods. We have applied the procedure to time series observed from GlobColour data related to Chlorophyll type-a concentrations and KD-490 levels in order to identify homogeneous areas in the Gulf of Gabes with respect to the temporal behavior of these water indicators \textcolor{black}{by means of a seasonal modulation model}. We defined clusters that are similar by different combination of quantiles of the two indicators. It is important to note that as different choice of quantiles implies changes in the clustering, this method may be particularly suitable for defining areas at different risk when considering two indicators at different quantile levels. More particularly, important features are of absolute interest in environmental sciences and ecology \citep{schmidt2012estimating}.
\textcolor{black}{In addition, the use of a model matrix based on a seasonal modulation model is evoked by the periodic behaviour of time series in our environmental application; however other flexible specifications can also be adopted.}

We note that both Chl-a concentrations and KD-490 levels are affected by several spatially varying factors. \textcolor{black}{Potential consequences of non incorporating in the model information related to the spatial domain may result from misclassification to  lower predictive ability; the spatial dependence can be particularly helpful in presence of high percentage of missing data. In this case potential future development of the modelling approach   can be the incorporation of  spatial dependence among the probabilities of membership as in   \cite{jiang2012clustering} or  \citet{Gaetan:2017}}.
Another possible extension is to perform clustering at multiple quantiles instead of fixing the levels of quantiles. However, caution has to be taken in this case to avoid the issue of crossing quantiles.

\bibliographystyle{natbib}

\begin{thebibliography}{}

\bibitem[Alikas {\em et~al.}(2015)]{Alikas:2015}
Alikas, K., Kangro, K., Randoja, R., Philipson, P., Asuk{\"u}ll, E., Pisek, J.,
  and Reinart, A. (2015).
\newblock Satellite-based products for monitoring optically complex inland
  waters in support of {EU} {W}ater {F}ramework {D}irective.
\newblock {\em International Journal of Remote Sensing}, {\bf 36}, 4446--4468.

\bibitem[Aloulou {\em et~al.}(2012)]{Aloulou:2012}
Aloulou, F., EllEuch, B., and Kallel, M. (2012).
\newblock Benthic foraminiferal assemblages as pollution proxies in the
  northern coast of {G}abes {G}ulf, {T}unisia.
\newblock {\em Environmental Monitoring and Assessment}, {\bf 184}, 777--795.

\bibitem[Alvera-Azc{\'a}rate {\em et~al.}(2012)]{alvera2012outlier}
Alvera-Azc{\'a}rate, A., Sirjacobs, D., Barth, A., and Beckers, J.-M. (2012).
\newblock Outlier detection in satellite data using spatial coherence.
\newblock {\em Remote Sensing of Environment}, {\bf 119}, 84--91.

\bibitem[Ayadi {\em et~al.}(2015)]{ayadi:2015}
Ayadi, N., Aloulou, F., and Bouzid, J. (2015).
\newblock Assessment of contaminated sediment by phosphate fertilizer
  industrial waste using pollution indices and statistical techniques in the
  {G}ulf of {G}abes ({T}unisia).
\newblock {\em Arabian Journal of Geosciences}, {\bf 8}, 1755--1767.

\bibitem[Barbosa {\em et~al.}(2011)]{Barbosa:2011}
Barbosa, S., Scotto, M., and Alonso, A. (2011).
\newblock Summarising changes in air temperature over {C}entral {E}urope by
  quantile regression and clustering.
\newblock {\em Natural Hazards and Earth System Sciences}, {\bf 11},
  3227--3233.

\bibitem[Barnard {\em et~al.}(2000)]{barnard:mcculloch:meng:2000}
Barnard, J., McCulloch, R., and Meng, X.-L. (2000).
\newblock Modeling covariance matrices in terms of standard deviations and
  correlations, with application to shrinkage.
\newblock {\em Statistica Sinica}, {\bf 10}, 1281--1311.

\bibitem[Benoit and Van~den Poel(2012)]{Benoit2012}
Benoit, D.~F. and Van~den Poel, D. (2012).
\newblock Binary quantile regression: a {B}ayesian approach based on the
  asymmetric {L}aplace distribution.
\newblock {\em Journal of Applied Econometrics}, {\bf 27}, 1174--1188.

\bibitem[Benoit {\em et~al.}(2013)]{Benoit2013}
Benoit, D.~F., Alhamzawi, R., and Yu, K. (2013).
\newblock {B}ayesian lasso binary quantile regression.
\newblock {\em Computational Statistics}, {\bf 28}, 2861--2873.

\bibitem[Cazelles {\em et~al.}(2008)]{Cazelles:2008}
Cazelles, B., Chavez, M., Berteaux, D., M{\'e}nard, F., Vik, J.~O., Jenouvrier,
  S., and Stenseth, N.~C. (2008).
\newblock Wavelet analysis of ecological time series.
\newblock {\em Oecologia}, {\bf 156}, 287--304.

\bibitem[Celeux {\em et~al.}(2006)]{celeux2006deviance}
Celeux, G., Forbes, F., Robert, C.~P., and Titterington, D.~M. (2006).
\newblock Deviance information criteria for missing data models.
\newblock {\em Bayesian Analysis}, {\bf 1}, 651--673.

\bibitem[Dabuleviciene {\em et~al.}(2020)]{Dabuleviciene:2020}
Dabuleviciene, T., Vaiciute, D., and Kozlov, I.~E. (2020).
\newblock Chlorophyll-a variability during upwelling events in the
  {South-Eastern} {Baltic} {Sea} and in the {Curonian} {Lagoon} from satellite
  observations.
\newblock {\em Remote Sensing}, {\bf 12}, 3661.

\bibitem[Directive {\em et~al.}(2000)]{directive2000european}
Directive, E. W.~F. {\em et~al.} (2000).
\newblock The {European} parliament and of the council.
\newblock {\em Water Framework Directive (2000/60/EC), OJL}, {\bf 327}, 1--73.

\bibitem[Downton(1970)]{Downton1970}
Downton, F. (1970).
\newblock Bivariate exponential distributions in reliability theory.
\newblock {\em Journal of the Royal Statistical Society: Series B}, {\bf 32},
  408--417.

\bibitem[Eilers {\em et~al.}(2008)]{eilers2008modulation}
Eilers, P.~H., Gampe, J., Marx, B.~D., and Rau, R. (2008).
\newblock Modulation models for seasonal time series and incidence tables.
\newblock {\em Statistics in Medicine}, {\bf 27}, 3430--3441.

\bibitem[El~Kateb {\em et~al.}(2016)]{el:2016}
El~Kateb, A., Stalder, C., Neururer, C., Pisapia, C., and Spezzaferri, S.
  (2016).
\newblock Correlation between pollution and decline of {S}cleractinian
  {C}ladocora {C}aespitosa ({L}innaeus, 1758) in the {G}ulf of {G}abes.
\newblock {\em Heliyon}, {\bf 2}, e00195.

\bibitem[El~Zrelli {\em et~al.}(2017)]{el:2017}
El~Zrelli, R., Courjault-Rad{\'e}, P., Rabaoui, L., Daghbouj, N., Mansour, L.,
  Balti, R., Castet, S., Attia, F., Michel, S., and Bejaoui, N. (2017).
\newblock Biomonitoring of coastal pollution in the {G}ulf of {G}abes (se,
  {T}unisia): use of {P}osidonia oceanica seagrass as a bioindicator and its
  mat as an archive of coastal metallic contamination.
\newblock {\em Environmental Science and Pollution Research}, {\bf 24},
  22214--22225.

\bibitem[El~Zrelli {\em et~al.}(2018)]{el:2018}
El~Zrelli, R., Rabaoui, L., Alaya, M.~B., Daghbouj, N., Castet, S., Besson, P.,
  Michel, S., Bejaoui, N., and Courjault-Rad{\'e}, P. (2018).
\newblock Seawater quality assessment and identification of pollution sources
  along the central coastal area of {G}abes {G}ulf (se {T}unisia): evidence of
  industrial impact and implications for marine environment protection.
\newblock {\em Marine Pollution Bulletin}, {\bf 127}, 445--452.

\bibitem[Finazzi {\em et~al.}(2015)]{Finazzi:2015}
Finazzi, F., Haggarty, R., Miller, C., Scott, M., and Fasso, A. (2015).
\newblock A comparison of clustering approaches for the study of the temporal
  coherence of multiple time series.
\newblock {\em Stochastic Environmental Research and Risk Assessment}, {\bf
  29}, 463--475.

\bibitem[Fourati {\em et~al.}(2018)]{Fourati:2018}
Fourati, R., Tedetti, M., Guigue, C., Goutx, M., Zaghden, H., Sayadi, S., and
  Elleuch, B. (2018).
\newblock Natural and anthropogenic particulate-bound aliphatic and polycyclic
  aromatic hydrocarbons in surface waters of the {G}ulf of {G}ab{\`e}s
  ({T}unisia, southern {M}editerranean sea).
\newblock {\em Environmental Science and Pollution Research}, {\bf 25},
  2476--2494.

\bibitem[Gaetan {\em et~al.}(2016)]{gaetan2016clustering}
Gaetan, C., Girardi, P., Pastres, R., Mangin, A., {\em et~al.} (2016).
\newblock Clustering chlorophyll-a satellite data using quantiles.
\newblock {\em Annals of Applied Statistics}, {\bf 10}, 964--988.

\bibitem[Gaetan {\em et~al.}(2017)]{Gaetan:2017}
Gaetan, C., Girardi, P., and Pastres, R. (2017).
\newblock Spatial clustering of curves with an application of satellite data.
\newblock {\em Spatial Statistics}, {\bf 20}, 110--124.

\bibitem[Giraldo {\em et~al.}(2012)]{Giraldo:2012}
Giraldo, R., Delicado, P., and Mateu, J. (2012).
\newblock Hierarchical clustering of spatially correlated functional data.
\newblock {\em Statistica Neerlandica}, {\bf 66}, 403--421.

\bibitem[Gr\"un and Leisch(2008)]{flexmix}
Gr\"un, B. and Leisch, F. (2008).
\newblock {FlexMix} version 2: Finite mixtures with concomitant variables and
  varying and constant parameters.
\newblock {\em Journal of Statistical Software}, {\bf 28}, 1--35.

\bibitem[Haggarty {\em et~al.}(2015)]{Haggarty:2015}
Haggarty, R., Miller, C., and Scott, E. (2015).
\newblock Spatially weighted functional clustering of river network data.
\newblock {\em Journal of the Royal Statistical Society. Series C, Applied
  Statistics}, {\bf 64}, 491--506.

\bibitem[Hamza and El~Abed(1994)]{hamza:1994}
Hamza, A. and El~Abed, A. (1994).
\newblock Les eaux color{\'e}es dans le golfe de {G}ab{\`e}s: bilan de six ans
  de surveillance (1989-1994).
\newblock {\em Bulletin de l'Institut National des Sciences et Technologies de
  la Mer}, {\bf 21}, 66--72.

\bibitem[Hubert and Arabie(1985)]{Hubert1985}
Hubert, L. and Arabie, P. (1985).
\newblock Comparing partitions.
\newblock {\em Journal of Classification}, {\bf 2}, 193--218.

\bibitem[Hyndman {\em et~al.}(2020)]{tsfeatures}
Hyndman, R., Kang, Y., Montero-Manso, P., Talagala, T., Wang, E., Yang, Y., and
  O'Hara-Wild, M. (2020).
\newblock {\em tsfeatures: Time Series Feature Extraction}.
\newblock R package version 1.0.2.

\bibitem[Jiang and Serban(2012)]{jiang2012clustering}
Jiang, H. and Serban, N. (2012).
\newblock Clustering random curves under spatial interdependence with
  application to service accessibility.
\newblock {\em Technometrics}, {\bf 54}, 108--119.

\bibitem[Jorgensen(1982)]{Jorgensen1982}
Jorgensen, B. (1982).
\newblock {\em Statistical {P}roperties of the {G}eneralized {I}nverse
  {G}aussian {D}istribution}.
\newblock Springer-Verlag, New York.

\bibitem[Katlane {\em et~al.}(2012)]{Katlane2012}
Katlane, R., DUPOUY, C., and Zargouni, F. (2012).
\newblock {Chlorophyll and turbidity concentrations deduced from MODIS as an
  index of water quality of the Gulf of Gabes in 2009}.
\newblock In AUF, editor, {\em {T{\'e}l{\'e}d{\'e}tection 11, 1}},
  T{\'e}l{\'e}d{\'e}tection, pages 265--273. CNRS \& Campus Spatial Univ. Paris
  Diderot VII.

\bibitem[Koenker and Bassett(1978)]{koenker1978regression}
Koenker, R. and Bassett, G. (1978).
\newblock Regression quantiles.
\newblock {\em Econometrica}, {\bf 46}, 33--50.

\bibitem[Koenker and Machado(1999)]{koenker1999goodness}
Koenker, R. and Machado, J.~A. (1999).
\newblock Goodness of fit and related inference processes for quantile
  regression.
\newblock {\em Journal of the American Statistical Association}, {\bf 94},
  1296--1310.

\bibitem[Kotz {\em et~al.}(2000)]{kotz:balakrishnan:johnson}
Kotz, S., N., B., and Johnson, N. (2000).
\newblock {\em Continuous Multivariate Distributions. Volume 1: Models and
  Applications}.
\newblock Wiley, New York.

\bibitem[Kotz {\em et~al.}(2001)]{Kotz2001}
Kotz, S., Kozubowski, T., and Podgorski, K. (2001).
\newblock {\em The {L}aplace Distribution and Generalizations: A Revisit with
  Applications to Communications, Economics, Engineering, and Finance}.
\newblock Springer, New York.

\bibitem[Li {\em et~al.}(2016)]{li2016bivariate}
Li, H., Deng, X., Dolloff, C., and Smith, E. (2016).
\newblock Bivariate functional data clustering: grouping streams based on a
  varying coefficient model of the stream water and air temperature
  relationship.
\newblock {\em Environmetrics}, {\bf 27}, 15--26.

\bibitem[Liechty {\em et~al.}(2004)]{liechty2004}
Liechty, J.~C., Liechty, M.~W., and M{\"u}ller, P. (2004).
\newblock Bayesian correlation estimation.
\newblock {\em Biometrika}, {\bf 91}, 1--14.

\bibitem[McLachlan and Peel(2000)]{mclachlan:peel:2000}
McLachlan, G.~J. and Peel, D. (2000).
\newblock {\em Finite {M}ixture {M}odels}.
\newblock Wiley, New York.

\bibitem[Monteiro {\em et~al.}(2012)]{Monteiro:2012}
Monteiro, A., Carvalho, A., Ribeiro, I., Scotto, M., Barbosa, S., Alonso, A.,
  Baldasano, J., Pay, M., Miranda, A., and Borrego, C. (2012).
\newblock Trends in ozone concentrations in the {I}berian {P}eninsula by
  quantile regression and clustering.
\newblock {\em Atmospheric {E}nvironment}, {\bf 56}, 184--193.

\bibitem[Pardo {\em et~al.}(2012)]{Pardo:2012}
Pardo, I., G{\'o}mez-Rodr{\'\i}guez, C., Wasson, J.-G., Owen, R., van~de Bund,
  W., Kelly, M., Bennett, C., Birk, S., Buffagni, A., Erba, S., {\em et~al.}
  (2012).
\newblock The {European} reference condition concept: a scientific and
  technical approach to identify minimally-impacted river ecosystems.
\newblock {\em Science of the Total Environment}, {\bf 420}, 33--42.

\bibitem[Petrella and Raponi(2019)]{Petrella2019a}
Petrella, L. and Raponi, V. (2019).
\newblock Joint estimation of conditional quantiles in multivariate linear
  regression models with an application to financial distress.
\newblock {\em Journal of Multivariate Analysis}, {\bf 173}, 70--84.

\bibitem[Poik{\=a}ne {\em et~al.}(2010)]{poikane2010defining}
Poik{\=a}ne, S., Alves, M.~H., Argillier, C., Van~den Berg, M., Buzzi, F.,
  Hoehn, E., De~Hoyos, C., Karottki, I., Laplace-Treyture, C., Solheim, A.~L.,
  {\em et~al.} (2010).
\newblock Defining chlorophyll-a reference conditions in {European} lakes.
\newblock {\em Environmental Management}, {\bf 45}, 1286--1298.

\bibitem[Rabaoui {\em et~al.}(2013)]{Rabaoui:2013}
Rabaoui, L., Balti, R., Zrelli, R., and Tlig-Zouari, S. (2013).
\newblock Assessment of heavy metals pollution in the {G}ulf of {G}abes
  ({T}unisia) using four mollusk species.
\newblock {\em {M}editerranean Marine Science}, {\bf 15}, 45--58.

\bibitem[Robert and Casella(2004)]{Robert2004}
Robert, C.~P. and Casella, G. (2004).
\newblock {\em Monte Carlo Statistical Methods}.
\newblock Springer, New York.

\bibitem[Schmidt {\em et~al.}(2012)]{schmidt2012estimating}
Schmidt, T.~S., Clements, W.~H., and Cade, B.~S. (2012).
\newblock Estimating risks to aquatic life using quantile regression.
\newblock {\em Freshwater Science}, {\bf 31}, 709--723.

\bibitem[Shi {\em et~al.}(2013)]{shi2013remote}
Shi, K., Li, Y., Li, L., Lu, H., Song, K., Liu, Z., Xu, Y., and Li, Z. (2013).
\newblock Remote chlorophyll-a estimates for inland waters based on a
  cluster-based classification.
\newblock {\em Science of the Total Environment}, {\bf 444}, 1--15.

\bibitem[Sottile and Adelfio(2019)]{Sottile:2019}
Sottile, G. and Adelfio, G. (2019).
\newblock Clusters of effects curves in quantile regression models.
\newblock {\em Computational Statistics}, {\bf 34}, 551--569.

\bibitem[Spiegelhalter {\em et~al.}(2002)]{spiegelhalter2002bayesian}
Spiegelhalter, D.~J., Best, N.~G., Carlin, B.~P., and Van Der~Linde, A. (2002).
\newblock Bayesian measures of model complexity and fit.
\newblock {\em Journal of the Royal Statistical Society: Series B (Statistical
  Methodology)}, {\bf 64}, 583--639.

\bibitem[Stafoggia {\em et~al.}(2017)]{Stafoggia:2017}
Stafoggia, M., Schwartz, J., Badaloni, C., Bellander, T., Alessandrini, E.,
  Cattani, G., De'Donato, F., Gaeta, A., Leone, G., Lyapustin, A., {\em et~al.}
  (2017).
\newblock Estimation of daily {PM10} concentrations in {I}taly (2006--2012)
  using finely resolved satellite data, land use variables and meteorology.
\newblock {\em Environment International}, {\bf 99}, 234--244.

\bibitem[Varin and Vidoni(2005)]{varin2005}
Varin, C. and Vidoni, P. (2005).
\newblock A note on composite likelihood inference and model selection.
\newblock {\em Biometrika}, {\bf 92}, 519--528.

\bibitem[Waldmann and Kneib(2015)]{Waldmann2015}
Waldmann, E. and Kneib, T. (2015).
\newblock {B}ayesian bivariate quantile regression.
\newblock {\em Statistical Modelling}, {\bf 15}, 326--344.

\bibitem[Wang {\em et~al.}(2006)]{Wang2006}
Wang, X., Smith, K., and Hyndman, R. (2006).
\newblock Characteristic-based clustering for time series data.
\newblock {\em Data Mining and Knowledge Discovery}, {\bf 13}, 335--364.

\bibitem[Yang {\em et~al.}(2020)]{Yang:2020}
Yang, C., Ye, H., and Tang, S. (2020).
\newblock Seasonal variability of diffuse attenuation coefficient in the
  {P}earl river estuary from long-term remote sensing imagery.
\newblock {\em Remote Sensing}, {\bf 12}, 2269.

\bibitem[Yu and Moyeed(2001)]{yu2001bayesian}
Yu, K. and Moyeed, R.~A. (2001).
\newblock {B}ayesian quantile regression.
\newblock {\em Statistics \& Probability Letters}, {\bf 54}, 437--447.

\bibitem[Zaghden {\em et~al.}(2014)]{Zaghden:2014}
Zaghden, H., Kallel, M., Elleuch, B., Oudot, J., Saliot, A., and Sayadi, S.
  (2014).
\newblock Evaluation of hydrocarbon pollution in marine sediments of {S}fax
  coastal areas from the {G}abes {G}ulf of {T}unisia, {M}editerranean {S}ea.
\newblock {\em Environmental Earth Sciences}, {\bf 72}, 1073--1082.

\bibitem[Zhang {\em et~al.}(2019)]{Zhang:2019}
Zhang, Y., Wang, H.~J., and Zhu, Z. (2019).
\newblock Quantile-regression-based clustering for panel data.
\newblock {\em Journal of Econometrics}, {\bf 213}, 54--67.

\end{thebibliography}

\newpage
\begin{center}
\section*{Supplementary material for\\ ``Clustering of bivariate satellite time series: a  quantile  approach''}	
\end{center}
\renewcommand{\thetable}{S.\arabic{table}}
\setcounter{table}{0}
\renewcommand{\thefigure}{S.\arabic{figure}}
\setcounter{figure}{0}

\renewcommand{\theequation}{S.\arabic{equation}}
\setcounter{equation}{0}
\setcounter{section}{0}
We  observe the set $\mathcal{Y} = \{\boldsymbol{y}_i , i = 1, \ldots, n\}$ of $n$ vectors  of independent observations, with 
$\boldsymbol{y}_i = (\tilde{y}_{i1}',\ldots, \tilde{y}_{iT}')'$, and
$\tilde{y}_{it} = (y_{i1t},y_{i2t},\ldots y_{iqt})'$, $t=1,\ldots,T$
from the hierarchical model 

\noindent\textbf{the observation level}
\begin{equation*}
	\tilde{y}_{it}|c_i,\tilde{w}_{it}\sim\mathcal{N}_q(X_{t}'\tilde\beta_{c_i}+ D_{c_i}\Theta \tilde{w}_{it},{W}_{it}D_{c_i}\Sigma_{c_i}D_{c_i}{W}_{it}), \qquad i=1,\ldots,n,\quad t=1,\ldots,T 
\end{equation*}
conditionally independent distributed, with $\Sigma_{c_i}=\Omega R(\tilde\phi_{c_i})\Omega$;

\noindent \textbf{the latent process level}
\begin{equation*}
	\tilde{w}_{it}|c_i\sim e(w,\gamma_{c_i}) \quad i=1,\ldots,n,\quad t=1,\ldots,T
\end{equation*}
conditionally independent distributed
\begin{equation*}
	c_i \sim \operatorname{Multinomial}(1,{\alpha}), \quad i=1,\ldots,n
\end{equation*}
independent distributed.

The prior distribution for the 
 the model parameters $\boldsymbol{\psi}=(
\alpha',\boldsymbol{\beta}',\boldsymbol{\gamma}',\boldsymbol{\sigma}',\boldsymbol{\phi}')'$ 
with $\alpha=(\alpha_1,\ldots,\alpha_K)'$, $\boldsymbol{\beta}=(\tilde\beta_1',\ldots,\tilde\beta_K')'$, 
$\boldsymbol{\gamma}=(\gamma_1',\ldots,\gamma_K')'$,
$\boldsymbol{\sigma}=(\tilde\sigma_1',\ldots,\tilde\sigma_K')'$ and 
$\boldsymbol{\phi}=(\tilde\phi_1',\ldots,\tilde\phi_K')'$.
are given by

\noindent $	 \pi(\boldsymbol{\beta})=\prod_{k=1}^K	\pi(\tilde\beta_{k})\propto \prod_{k=1}^K
|P_k|^{-1/2}\exp\left\{-\frac 12
(\tilde{\beta}_{k}-b_k)'P_k^{-1}(\tilde{\beta}_{k}-b_k)\right\}
$;

\noindent $\pi(\boldsymbol{\sigma})=\prod_{k=1}^K	\pi(\tilde\sigma_{k})\propto \prod_{k=1}^K
\prod_{j=1}^q {\sigma^2_{kj}}^{-s_{k}-1}\exp\left(-{d_{k}}/{\sigma^2_{kj}}\right)
$;

\noindent  $\pi(\boldsymbol{\phi})=\prod_{k=1}^K	\pi(\tilde\phi_{k})$ and  $\pi(\boldsymbol{\gamma})=\prod_{k=1}^K	\pi(\gamma_{k})$. More details on the analytical form of $\pi(\tilde\phi_{k})$ and $\pi(\gamma_{k})$ will be given later.

\noindent $
\pi({\alpha})=\frac{\varGamma(a_{1}+...+a_{K})}{\varGamma(a_{1})\cdot...\cdot\varGamma(a_{K})}\prod_{k=1}^K\alpha_{k}^{a_{k}-1}
$ 	 with parameters $a_1, \ldots, a_K > 0$

Finally we denote  the set $\mathcal{W}=(\boldsymbol{w}_1',\ldots, \boldsymbol{w}_n')$, with latent vectors
$\boldsymbol{w}_i = (\tilde{w}_{i1}',\ldots, \tilde{w}_{iT}')'$, 
$\tilde{w}_{it} = (w_{i1t},y_{i2t},\ldots w_{iqt})'$, $t=1,\ldots,T$
 the (latent) vector of the cluster memberships $\mathcal{C} = (c_1 ,\ldots, c_n )$ 

We now proceed to deriving the conditional  distribution of the model parameters including also the conditional  for the latent quantities $\mathcal{W}$
and $\mathcal{C}$.

Applying the Bayes theorem we obtain the  posterior distribution as
\begin{eqnarray}
	\pi({\alpha},\boldsymbol{\beta},\boldsymbol{\gamma},\boldsymbol{\sigma},\boldsymbol{\phi},\mathcal{C},\mathcal{W}|\mathcal{Y})&\propto&\prod_{i=1}^{n} 
	f(\boldsymbol{y}_i|{\alpha},\tilde{\beta}_{c_i},\tilde{\sigma}_{c_i}, \tilde{\phi}_{c_i}, \boldsymbol{w}_i, c_i)\times\nonumber\\
	&& \prod_{i=1}^{n} 
		  f(\boldsymbol{w}_{i}|\gamma_{c_{i}})
	\times f(c_{i}|{\alpha})\times \nonumber\\
	&&\prod_{k=1}^K\pi(\tilde\beta_{k})\times \pi(\tilde\sigma_{k})\times\pi(\tilde\phi_{k})\times \pi(\gamma_{k}) \times\pi({\alpha}) \label{eq3.13}
\end{eqnarray}
where 

\noindent $\begin{array}{ll}
		f(\boldsymbol{y}_i|{\alpha},\tilde{\beta}_{c_i},\tilde{\sigma}_{c_i}, \tilde{\phi}_{c_i}, \boldsymbol{w}_i, c_i)\propto&\prod_{t=1}^T
|{W}_{it}D_{c_i}\Sigma_{c_i}D_{c_i}{W}_{it}|^{-1/2}\exp\left\{-\frac 12
(\tilde{y}_{it}-X_{t}'\tilde\beta_{c_i}- D_{c_i}\Theta
\tilde{w}_{it})'\times\right.\\
&[{W}_{it}D_{c_i}\Sigma_{c_i}D_{c_i}{W}_{it}]^{-1}(\tilde{y}_{it}-X_{t}'\tilde\beta_{c_i}- D_{c_i}\Theta
\tilde{w}_{it})\left.\right\};
\end{array}
$

\subsubsection*{Full conditional  for $\alpha$}

To derive the  posterior density of the probability vector $ {\alpha} $, we first note that 
$	\pi({\alpha},|\boldsymbol{\beta},\boldsymbol{\gamma},\boldsymbol{\sigma},\boldsymbol{\phi},\mathcal{C},\mathcal{W},\mathcal{Y})\propto \pi(\alpha|\mathcal{C})$ and then
\begin{equation*}
\pi({\alpha}|\boldsymbol{\beta},\boldsymbol{\sigma},\boldsymbol{\gamma},\boldsymbol{\phi},\mathcal{C},\mathcal{W},\mathcal{Y})	\propto \prod_{i=1}^{n} 
 f(c_{i}|{\alpha})\pi({\alpha})\propto \prod_{k=1}^{K}\alpha_{k}^{a_{k}+{\sum_{i=1}^{n}I(c_{i}=k)}-1} \label{eq3.16}
\end{equation*}
Therefore  $\pi({\alpha}|\boldsymbol{\beta},\boldsymbol{\gamma},\boldsymbol{\sigma},\boldsymbol{\phi},\mathcal{C},\mathcal{W},\mathcal{Y})$ is the density of a Dirichlet distribution with parameters ${a_{k}+{\sum_{i=1}^{n}I(c_{i}=k)}-1}$, $k=1,\ldots,K$.
\subsubsection*{Full conditional  for $\boldsymbol{\beta}$}
\noindent
We have
\begin{eqnarray*}
	\pi(\boldsymbol{\beta}|\alpha,\boldsymbol{\gamma},\boldsymbol{\sigma},\mathcal{C},\mathcal{W},\mathcal{Y})
	\propto\prod_{k=1}^K
		\pi(\tilde\beta_k|{\tilde\sigma}_k,\tilde\phi_k,\mathcal{C},\mathcal{W},\mathcal{Y})
\end{eqnarray*}
where
\begin{eqnarray*}
	\pi(\tilde\beta_k|{\tilde\sigma}_k,\tilde\phi_k,\mathcal{C},\mathcal{W},\mathcal{Y})
	&\propto& \exp\left\{-\frac{1}{2}(\tilde\beta_{k}-b_k)'P_k^{-1}(\tilde\beta_{k}-b_k)\right\}\times\\
	&&	\prod_{i=1}^{n}
	\left[ \exp\left\{-\frac{1}{2}\sum_{t=1}^{T}\left(\tilde u_{it}-X_{t}'\tilde\beta_{c_i}\right)'S_{c_i,t}^{-1}\left(\tilde u_{it}-X_{t}'\tilde\beta_{c_i}\right)\right\}	\right]^{I(c_i=k)},
\\
\end{eqnarray*}	
with $\tilde{u}_{it}=(u_{i1t},\ldots,u_{iqt})'=\tilde{y}_{it}-D_{c_i}\Theta\,
\tilde{w}_{it}$ and 
$S_{c_i,t}={W}_{it}D_{c_i}\Sigma_{c_i}D_{c_i}{W}_{it}$

The previous formula can be further elaborated, namely
\begin{eqnarray*}
		\pi(\tilde\beta_k|{\tilde\sigma}_k,\tilde\phi_k,\mathcal{C},\mathcal{W},\mathcal{Y})
	&\propto& 	\exp\left\{-\frac{1}{2}\left[\tilde\beta_{k}'\left(P_k^{-1}+\sum_{i=1}^{n}I(c_i=k)\,\tilde{X}'S_{c_i}^{-1}\tilde{X} \right)\tilde\beta_{k}-\right.\right.\\
	&&\left.\left. 2\tilde\beta_{k}'\left(P_k^{-1}b_k+\sum_{i=1}^{n}I(c_i=k)\,\tilde{X}'S_{c_i}^{-1}\boldsymbol{u}_{i}\right)\right]\right\}
\end{eqnarray*}	
where $ S_{c_i} $ is a block diagonal matrix with entries $S_{c_i,t}$, $t=1,\ldots,T$, $ \tilde{X}=[X_1,\ldots,X_T]' $ is the matrix of covariates and $ \boldsymbol{u}_{i}=(\tilde{u}_{i1}',...,\tilde{u}_{iT}')' $.

Therefore $\pi(\tilde\beta_k|{\tilde\sigma}_k,\tilde\phi_k,\mathcal{C},\mathcal{W},\mathcal{Y})$
 is  the  density, up to a normalizing constant, of a multivariate Gaussian vector  with vector mean 

$$
\bar{m}_{k}=\bar{S}_k^{-1}\left(P_k^{-1}b_k+\sum_{i=1}^{n}I(c_i=k)\,X'S_{c_i}^{-1}\boldsymbol{u}_{i}\right)
$$
and covariance matrix
$$
\bar{S}_k=\left(P_k^{-1}+\sum_{i=1}^{n}I(c_i=k)\,X'S_{c_i}^{-1}X\right)^{-1}.
$$

\subsubsection*{Full conditional  for $\boldsymbol{\sigma}$}

We have $$	\pi(\boldsymbol{\sigma}|\alpha,\boldsymbol{\beta},\boldsymbol{\gamma},\boldsymbol{\phi},\mathcal{C},\mathcal{W},\mathcal{Y}) 
	\propto\prod_{k=1}^K \pi(\tilde\sigma_k|\tilde{\beta}_k,\tilde\phi_k,\mathcal{C},\mathcal{W},\mathcal{Y})
$$	
where

\begin{eqnarray*}
	\pi(\tilde\sigma_k|\tilde{\beta}_k,\tilde\phi_k,\mathcal{C},\mathcal{W},\mathcal{Y})
	&\propto&\prod_{j=1}^q [\sigma_{kj}^2]^{-(s_{j}+1)} \exp\left(-\frac{d_{j}}{\sigma_{kj}^2}\right)\times 
	\\
	&&
		\prod_{i=1}^{n}\left\{|\tilde{S}_{c_i}|^{-1/2}
	 \exp\left[-\frac{1}{2}\sum_{t=1}^{T}\left(\tilde u_{it}-X_{t}'\tilde\beta_{c_i}\right)'S_{c_i,t}^{-1}\left(\tilde u_{it}-X_{t}'\tilde\beta_{c_i}\right)\right]	\right\}^{I(c_i=k)},
\\	
\end{eqnarray*}
However, it is difficult to calculate the normalizing constant for this density and sample from the conditional distribution.
 As a result we introduce a Metropolis-Hastings (M-H) step in our simulation algorithm. The proposals in the M-H step are the following.
 
 We pretend that $\tilde\phi_k=0$ and in this case 
$$
 	\pi(\tilde\sigma_k|\tilde{\beta}_k,0,\mathcal{C},\mathcal{W},\mathcal{Y})
 	 	= 	\prod_{j=1}^q\pi(\sigma_{kj}|\tilde{\beta}_k,0,\mathcal{C},\mathcal{W},\mathcal{Y}) 
 	 	$$
with 	 	
  \begin{eqnarray*}	 	
\pi(\sigma_{kj}|\tilde{\beta}_k,0,\mathcal{C},\mathcal{W},\mathcal{Y}) 	 	&\propto& [\sigma_{kj}^2]^{-(s_{j}+1)} \exp\left(-\frac{d_{j}}{\sigma_{kj}^2}\right)\times\\ 
 		&&\prod_{i=1}^{n}
	\left\{ (\sigma_{c_{i},j}^2)^{-{T}/{2}}\exp\left[-\frac{1}{2
		\sigma_{c_{i},j}^2\omega^{2}}\sum_{t=1}^{T}\frac{\left(u_{ijt}-X_{jt}'\beta_{c_{i}}\right)^{2}}{w_{ijt}}\right]\right.\times\\
	&&\left.
(\sigma_{c_{i,j}}^2
)^{-T}\exp\left(-\frac{1}{\sigma_{c_{i},j}^2}\sum_{t=1}^{T}w_{ijt}\right)\right\}^{I(c_i=k)}\nonumber\\
	&\propto& (\sigma_{kj}^2)^{-(2s_{j}+3n_{k}T)/2-1}\times \nonumber\\ 
	&&\exp\left
	\{-\frac{1}{\sigma_{kj}^2}\left[d_{j}+\sum_{i=1}^{n}\left(\sum_{t=1}^{T}w_{ijt}+\frac{1}{2\omega^{2}}\sum_{t=1}^{T}\frac{\left(u_{ijt}-X_{jt}'\beta_{c_{i}}\right)^{2}}{w_{ijt}}\right)I(c_i=k)\right]\right\}\nonumber\\
\end{eqnarray*}
Here $X_{jt}$ is the $j$-th row of the matrix $X_t$ and $n_{k}=\sum_{i=1}^nI(c_i=k)$, the number of vectors in $ \mathcal{Y} $ with membership $ k $.

Thus $\pi(\sigma_{kj}|\tilde{\beta}_k,0,\mathcal{C},\mathcal{W},\mathcal{Y})$  is the distribution of an inverse Gamma random variable, $\mathrm{InvGamma}(a_{kj}, b_{kj})$, with  shape  and scale parameter  
\begin{eqnarray}
	a_{kj}&=&(2s_{j}+3n_{k}T)/2 \nonumber\\
	b_{kj}&=&
	\left[d_{j}+\sum_{i=1}^{n}\left(\sum_{t=1}^{T}w_{ijt}+\frac{1}{2\omega^{2}}\sum_{t=1}^{T}\frac{\left(u_{ijt}-X_{jt}'\beta_{c_{i}}\right)^{2}}{w_{ijt}}\right)I(c_i=k)\right]\label{eq:InvGamma}
\end{eqnarray}
Note that random samples from inverse Gamma distribution can be drawn from a  Gamma distribution exploiting the fact that  if $G \sim \mathrm{Gamma}(a, 1/b)$  then $1/G \sim \mathrm{InvGamma}(a, b)$.

The proposal in  M-H	step is accomplished by sampling independent values  $ \sigma_{kj}\sim \mathrm{InvGamma}(a_{kj}, b_{kj}) $, $j=1,\ldots,q$    
where the shape  and scale parameters depend on the previous values of the chain. This proposal density is the full conditional of $ \tilde\sigma_k $  for $\tilde\phi=0$ and thus it leads to having higher acceptance rates for chain values with $\tilde\phi$ close to zero than in other cases.

\subsubsection*{Full conditional  for $\boldsymbol{\phi}$}
Using the same arguments  as for $\boldsymbol{\sigma}$ we have
$$	\pi(\boldsymbol{\phi}|\alpha,\boldsymbol{\beta},\boldsymbol{\gamma},\boldsymbol{\sigma},\mathcal{C},\mathcal{W},\mathcal{Y}) 
\propto\prod_{k=1}^K \pi(\tilde\phi_k|\tilde{\beta}_k,\tilde\sigma_k,\mathcal{C},\mathcal{W},\mathcal{Y})
$$	
with
$$
	\pi(\tilde\phi_k|\tilde{\beta}_k,\tilde\sigma_k,\mathcal{C},\mathcal{W},\mathcal{Y})
	\propto \pi(\tilde\phi_k) 	\prod_{i=1}^{n}\left\{|\tilde{S}_{c_i}|^{-1/2}
	\exp\left[-\frac{1}{2}\sum_{t=1}^{T}\left(\tilde u_{it}-X_{t}'\tilde\beta_{c_i}\right)'S_{c_i,t}^{-1}\left(\tilde u_{it}-X_{t}'\tilde\beta_{c_i}\right)\right]	\right\}^{I(c_i=k)},
$$
Here $\pi(\tilde\phi_k)$ is the prior distribution for $\tilde\phi_k$.
\citet{barnard:mcculloch:meng:2000} 
proposed two alternative prior models for $\tilde\phi_k$. One is the marginally uniform prior, in which
the marginal prior for each correlation  is a modified beta distribution over $[-1, 1]$; with an
appropriate choice of the beta parameters, this becomes a uniform marginal prior distribution. The other model for $\tilde\phi_k$ is called the jointly uniform prior.  \citet{liechty2004} discussed prior  uniformly distributed over all possible correlation matrices.
Under the bivariate case, $q=2$,
the choice is greatly simplified and we  choose a uniform prior, i.e.
$\phi_k\sim\mathcal{U}(-1,1)$ for every $k$.

Even with this simple choice, the conditional simulation requires a M-H step.
	The  independent proposal for $\phi_k$ is drawn from a uniform distribution ``centered'' around the current value in the chain, say $ \phi_k^* $, i.e.
$$\phi_k\sim U(\max\{-1,\phi_k^*-r\},\min\{1,\phi_k^*+r\})$$
with  $r >0$. In the simulation experiments we have seen that for a value like as $r= 0.1$ the MCMC algorithm performs very well.

\subsubsection*{Full conditional  for $\mathcal{C}$}
\noindent
We start by noting that 
$$f(\mathcal{C}|{\alpha},\boldsymbol{\beta},\boldsymbol{\gamma},\boldsymbol{\sigma},\mathcal{W},\mathcal{Y})\propto \prod_{i=1}^nf(c_{i}|
\alpha,\tilde\beta_{c_{i}},\tilde\sigma_{c_{i}},\phi_{c_{i}},\boldsymbol{y}_{i},\boldsymbol{w}_{i})$$

The conditional probability of the membership  of $c_i$ is given by
\begin{eqnarray}
	f(c_{i}|\boldsymbol{y}_{i},\boldsymbol{w}_{i},\beta_{c_{i}},\sigma_{c_{i}},\alpha)&=&\cfrac{\alpha_{c_i}		f(\boldsymbol{y}_i|{\alpha},\tilde{\beta}_{c_i},\tilde{\sigma}_{c_i}, \tilde{\phi}_{c_i}, \boldsymbol{w}_i, c_i)}{\sum_{k=1}^{K}\alpha_{k}
		f(\boldsymbol{y}_i|{\alpha},\tilde{\beta}_{k},\tilde{\sigma}_{k}, \tilde{\phi}_{k}, \boldsymbol{w}_i, k)} \label{eq3.17}
\end{eqnarray}

\subsubsection*{Full conditional  for $\mathcal{W}$}
We note  that 

$$	
	\pi(\mathcal{W}|\alpha,\boldsymbol{\beta},\boldsymbol{\gamma},\boldsymbol{\sigma},\boldsymbol{\phi},\mathcal{C},\mathcal{Y})\propto \prod_{i=1}^n
	f(\boldsymbol{w}_{i}|\tilde\beta_{c_{i}},\gamma_{c_{i}},\tilde\sigma_{c_{i}},\tilde\phi_{c_{i}},c_{i},\boldsymbol{y}_{i})=\prod_{i=1}^n\prod_{t=1}^T
	f(\tilde{w}_{it}|\tilde\beta_{c_{i}},\gamma_{c_{i}},\tilde\sigma_{c_{i}},
	\tilde\phi_{c_{i}},c_{i},\tilde{y}_{it})
$$
with
$$
f(\tilde{w}_{it}|\tilde\beta_{c_{i}},\gamma_{c_{i}},\tilde\sigma_{c_{i}},
\tilde\phi_{c_{i}},c_{i},\tilde{y}_{it})\propto e(\tilde{w}_{it};\gamma_{c_{i}},)
	|\tilde{S}_{c_i}|^{-1/2}
	\exp\left[-\frac{1}{2}\left(\tilde u_{it}-X_{t}'\tilde\beta_{c_i}\right)'S_{c_i,t}^{-1}\left(\tilde u_{it}-X_{t}'\tilde\beta_{c_i}\right)\right]\\
$$
Here  ${e}(\tilde{w}_{it};\gamma)$ is a multivariate density function with marginal unit exponential distributions and $\gamma$ is a generic dependence parameter.
The density $f(\tilde{w}_{it}|\tilde\beta_{c_{i}},\gamma_{c_{i}},\tilde\sigma_{c_{i}},
\tilde\phi_{c_{i}},c_{i},\tilde{y}_{it})$  seems not available in closed form for any reasonable choice of $e(\tilde{w}_{it};\gamma)$.
Once again we resort to a M-H step for simulating from it.

In case of independence of the components of $\tilde{w}_{it}$, conventionally identified with $\gamma_{k}=0$, for all $k$, we have
$e(\tilde{w}_{it};0)=\prod_{j=1}^q \exp(-w_{ijt})$. Moreover
we pretend that $\tilde\phi_k=0$ and in this case 
$$
f(\tilde{w}_{it}|\tilde\beta_{c_{i}},0,\tilde\sigma_{c_{i}},
0,c_{i},\tilde{y}_{it})
=\prod_{j=1}^q e({w}_{ijt}|\tilde\beta_{c_{i}},\sigma_{c_{i},j},c_{i},{y}_{ijt})
$$
where
\begin{eqnarray*}
	e(\tilde{w}_{ijt}|\tilde\beta_{c_{i}},\sigma_{c_{i},j},c_{i},y_{ijt})
	&\propto& w_{ijt}^{-{1}/{2}}\exp\left\{-\frac{1}{2\sigma_{c_{i},j}^2\omega^{2}w_{ijt}}\left(y_{ijt}-X_{jt}'\tilde\beta_{c_{i}}-\theta w_{ijt}\right)^{2}-\frac{w_{ijt}}{\sigma_{c_{i},j}^2}\right\}\\
	&\propto& w_{ijt}^{-{1}/{2}}\exp\left[-\frac{1}{2\sigma_{c_{i},j}^2\omega^{2}w_{ijt}}\left\{(y_{it}-X_{jt}'\tilde\beta_{c_{i}})^{2}+\theta^{2}w_{ijt}^{2}\right\}-\frac{w_{ijt}}{\sigma_{c_{i},j}^2}\right]\\
	&\propto& w_{ijt}^{-{1}/{2}}\exp\left\{-\frac{1}{2\sigma_{c_{i},j}^2\omega^{2}w_{ijt}}(y_{it}-X_{jt}'\tilde\beta_{c_{i}})^{2}-\frac{\theta^{2}w_{ijt}}{2\sigma_{c_{i},j}^2\omega^{2}}-\frac{w_{ijt}}{\sigma_{c_{i},j}^2}\right\}\\
	&\propto& w_{ijt}^{-{1}/{2}}\exp\left\{-\frac{1}{2\sigma_{c_{i},j}^2\omega^{2}w_{ijt}}(y_{it}-X_{jt}'\tilde\beta_{c_{i}})^{2}-\frac{w_{ijt}}{2\sigma_{c_{i},j}^2\omega^{2}}(\theta^{2}+2\omega^{2})\right\}\\
	&\propto& w_{ijt}^{-{1}/{2}}\exp\left\{-\frac{1}{2\sigma_{c_{i},j}^2\omega^{2}w_{ijt}}(y_{it}-X_{jt}'\tilde\beta_{c_{i}})^{2}-\frac{\omega^{2}w_{ijt}}{8\sigma_{c_{i},j}^2}\right\}\\
	&\propto& w_{ijt}^{-{1}/{2}}\exp\left\{-\frac{1}{2}\left(\frac{\omega^{2}w_{ijt}}{4\sigma_{c_{i},j}^2}+\frac{(y_{it}-X_{jt}'\tilde\beta_{c_{i}})^{2}}{\sigma_{c_{i},j}^2\omega^{2}w_{ijt}}\right)\right\}.
\end{eqnarray*}
This expression resembles a Generalized Inverse Gaussian distribution $ \text{GIG}(p,a,b) $ with the density
\begin{eqnarray*}
	h(w;a,b,p)&\propto& w^{(p-1)}\exp\left\{-\frac{1}{2}\left(aw+\frac{b}{w}\right)\right\}
\end{eqnarray*}
where $\displaystyle	a=\frac{\omega^{2}}{4\sigma_{c_{i},j}^2}$, $\displaystyle b=\frac{(y_{ijt}-X_{jt}'\tilde\beta_{c_{i}})^{2}}{\sigma_{c_{i},j}^2\omega^{2}}$
and $p=1/2$. 
\cite{Jorgensen1982} notes that if $W\sim\text{GIG}(-p,b,a)$ then $1/W\sim\text{GIG}(p,a,b) $ and further that the distribution  $\text{GIG}(-1/2,a,b)$ equals the Inverse Gaussian distribution 
\begin{eqnarray*}
	g(w|\mu,\lambda)&\propto& w^{-{3}/{2}}\exp\left(\frac{-\lambda(w-\mu)^{2}}{2\mu^{2}w}\right)\\
	&\propto& w^{-{3}/{2}}\exp\left(-\frac{\lambda w}{2\mu^{2}}-\frac{\lambda}{2w}\right)
\end{eqnarray*}
with $ a={\lambda}/{\mu^{2}} $, $ b=\lambda $ and $ p=-{1}/{2} $.

By recognizing these facts we can sample from the conditional posterior of $ w_{ijt} $ by drawing from the Inverse Gaussian distribution with
$\displaystyle\lambda=\frac{\omega^{2}}{4\sigma_{c_{i},j}^2}$ and $\displaystyle \frac{\lambda}{\mu^{2}}=\frac{(y_{ijt}-X_{jt}'\tilde\beta_{c_{i}})^{2}}{\sigma_{c_{i},j}^2\omega^{2}}$ that implies
$\displaystyle\mu=\frac{\omega^{2}}{2|y_{ijt}-X_{jt}'\tilde\beta_{c_{i}}|}$.

Threfore, by choosing  $	e(\tilde{w}_{ijt}|\tilde\beta_{c_{i}},\sigma_{c_{i},j},c_{i},y_{ijt})$ as proposal density  for every component ${w}_{ijt}$ of $ \tilde{w}_{it} $,  we shall propose a value $w$ such that $1/w$ is drawn from the density
$\displaystyle
\text{InvGauss}\left(\frac{\omega^{2}}{2|y_{ijt}-X_{jt}'\tilde\beta_{c_{i}}|}, \frac{\omega^{2}}{4\sigma_{c_{i},j}^2}\right) 
$.

\subsubsection*{Full conditional  for $\boldsymbol{\gamma}$}

We have $$	\pi(\boldsymbol{\gamma}|\alpha,\boldsymbol{\beta},\boldsymbol{\sigma},\boldsymbol{\phi},\mathcal{C},\mathcal{W},\mathcal{Y}) 
\propto\prod_{k=1}^K \pi(\gamma_k|\mathcal{C},\mathcal{W})
$$	
where
$$
\pi(\gamma_k|\mathcal{C},\mathcal{W})
	\propto\pi(\gamma_k)
	\prod_{i=1}^{n}\left\{
	\prod_{t=1}^{T}{e}(\tilde{w}_{it};\gamma_{c_i})\right\}^{I(c_i=k)}
$$

Here $\pi(\gamma_k)$ is the prior distribution for $\gamma_k$.
The density $\pi(\gamma_k|\mathcal{C},\mathcal{W})$
seems not available in closed form for any reasonable choice of $e(\tilde{w}_{it};\gamma)$.
Once again we resort to a M-H step for simulating from it.

\end{document}